\documentclass[a4paper,12pt]{article}
\usepackage{amssymb}
\usepackage{amsmath}
\usepackage{graphicx}
\usepackage{color}

\newcommand\nn{\nonumber}

\newcommand\beq{\begin{equation}}
\newcommand\eeq{\end{equation}}
\newcommand\bea{\begin{eqnarray} }
\newcommand\eea{\end{eqnarray} }

\begin{document}

\begin{titlepage}


\begin{center}

\bigskip

{\Large\bf{Novel Parity Violating Transport Coefficients in $2+1$ Dimensions from Holography}}

\vfill

Jiunn-Wei Chen$^{a\,b}$\,\footnote{\tt jwc@phys.ntu.edu.tw}, 
Shou-Huang Dai$^c$\,\footnote{\tt shdai.hep@gmail.com}, 
Nien-En Lee$^a$\,\footnote{\tt r99222006@ntu.edu.tw}, 
Debaprasad Maity$^{a\,b}$\,\footnote{\tt debu.imsc@gmail.com}

\bigskip

{\it $^a$Department of Physics and Center for Theoretical Sciences,\\ 
National Taiwan University, Taipei 10617, TAIWAN\\}
{\it $^b$Leung Center for Cosmology and Particle Astrophysics,\\
National Taiwan University, Taipei 10617, TAIWAN\\}
{\it $^c$Department of Physics, National Taiwan Normal University,\\
88, Sec.4, Ting-Chou Road, Taipei 11677, TAIWAN\\}
\end{center}

\vfill

\begin{abstract}
We construct a 3+1 dimensional holographic model dual to 
a parity violating hydrodynamic system in 2+1 dimensions. Our model contains gravitational and electrodynamic 
Chern-Simons terms coupled to a neutral pseudo scalar $\theta$, and a potential composed of quadratic and quartic terms in $\theta$. The background is a charged black brane. 
We study the hydrodynamics to first order in spacetime derivatives near the probe limit of the pseudo scalar, by extracting the transport coefficients from the scalar, vector, and tensor modes of bulk perturbations. We study two mechanisms for breaking the parity of the boundary fluid: the parity is either spontaneously broken by the nonzero vev of the dual pseudo scalar operator, or by the pseudo scalar source on the boundary. We discover some novel temperature-dependent behaviors of the transport coefficients. It would be interesting to find these behaviors being realized in the real world materials..
\end{abstract}

\vfill


\end{titlepage}

\setcounter{footnote}{0}

\section{Introduction} \label{sec:intro}

In the last decade, AdS/CFT correspondence has been demonstrated to be a useful tool to study the  strongly coupled gauge theories. One of the particular 
applications recently attracting significant 
attentions is holographic hydrodynamics \cite{son, Bhattacharyya:2008jc}. The hydrodynamics that 
is particularly well described by this approach so far is 
relativistic in nature. By applying this technique, various hydrodynamics transport coefficients of certain classes of strongly coupled conformal field theories have been computed subject to the underlying symmetry. According to this holography, the near horizon physics of the regular black hole in an 
asymptotically AdS spacetime is able to describe the long
wavelength dynamics of a strongly coupled finite temperature
field theory on the boundary. 

It is well known that the conventional hydrodynamics can be constructed from the underlying symmetry of the system without knowing much details about the microscopic degrees of freedom. In the hydrodynamic regime, a system can be completely described by a set of conservation equations, the relativistic generalization of the well-known 
Navier-Stokes equations. If the system has some dissipative 
parts in the conservation equations, their form are subject to the second law of thermodynamics. One interesting feature of the aforementioned holographic method \cite{Bhattacharyya:2008jc} is that it turns out to incorporate the constraint of positivity of entropy production, and successfully reproduces the relativistic hydrodynamic equations without referring to this constraint. It has been conjectured by an explicit example that the solutions of the relativistic hydrodynamic equations are in fact
in one to one correspondence with the regular long wavelength solutions of the bulk Einstein equations in 
AdS spacetime. Afterwards, the 
holographic technique had been applied to a wide range of 
strongly coupled systems dual to various kinds of asymptotic AdS black hole backgrounds \cite{van, siraz2, yarom, Banerjee:2008th}, and further generalized to describe the superfluid phase \cite{sirazherzog}.

Besides being an elegant complimentary approach towards deriving the hydrodynamic equations, the holographic approach provides us new transport coefficients which do not appear in the conventional phenomenological 
construction. This essentially inspires us to reconsider 
the whole approach of constructing the hydrodynamics from more general perspective. Positivity of the local entropy production is an important guiding principle for specifying the constitutive relations among the thermodynamic variables. It is first pointed out in \cite{son3} that, in order to accommodate those new terms in 
the stress-energy tensor and the global current, the entropy current has to be modified. Based on this, attempts have been made to construct the most general 
relativistic hydrodynamics order by order in derivative expansions, to incorporate various field theory anomaly- induced transport coefficients \cite{loga, Jensen:2012jy}. See also \cite{Jensen:2012jh} for the recent proposals of constructing hydrodynamics from the field theoretic perspective without referring to the entropy current.

The holographic approach can be applied to the systems with 
parity violation \cite{son2,Chen,yarom2,kimura}. Recently, parity violating transport in high energy theory has gained 
significant interests, due to the charge asymmetry observed in the heavy-ion collision experiments such as STAR \cite{star}, PHENIX \cite{phenix} and ALICE \cite{alice}. The parity violating chiral magnetic effect (CME) was proposed to explain the aforementioned 
observation and subsequently under intensive study during the past few years \cite{kharzeev}. On the other hand, the low energy transport properties of parity violating systems had long been the subject of interest in the condensed matter research. Anomalous Hall conductivity is one of such effects which had been measured experimentally, see \cite{nagaosa} for recent review. The thermal Hall conductivity in the presence of background magnetic fields is demonstrated to be a useful probe for the quasiparticle transport in the high temperature superconductors \cite{htcond}. There are also discussions on the anomaly-induced thermal Hall effect in topological insulators \cite{ryu}. A unified field theoretic model had been proposed for the anomalous electric, thermoelectric, and thermal Hall effects in the ferromagnetic materials \cite{onoda}. In hydrodynamics, the Hall viscosity, which is the parity violating analog of shear viscosity, is already studied in the field theoretic approach \cite{avron, hoyos}.

When the parity of the underlying theory of a system is broken, extra transport coefficients 
arise in the long wavelength limit. For the strongly interacting fluids, these new coefficients are difficult to evaluate from field theory perspective. One well-studied example is the Hall conductivity. From the field theory point of view, the Hall conductivity may originate from a bare Chern-Simons term in the fundamental theory, or the Chern-Simons term induced by radiative corrections. Even in the lattice computation which is very useful when the coupling is strong, there is a sign problem occurring to the Chern-Simons term after performing the Wick rotation. However, the AdS/CFT correspondence offers a powerful non-perturbative tool to calculate these new transport coefficients. In this paper, we use this method to compute all possible transport coefficients in first order derivative expansion of a holographic, strongly coupled, charged fluid in $2+1$ 
dimensions with parity violation in the hydrodynamic limit.

This paper generalizes our previous construction in  \cite{Chen} by considering Chern-Simons terms coming from the electromagnetic sector, $F\tilde{F}$, and the gravitational sector, $R\tilde{R}$, both coupled to a neutral pseudo scalar field. Our four dimensional gravitational background is a charged black brane in the probe limit of the pseudo scalar. The parity of the boundary fluid is broken by the operator dual to the bulk pseudo scalar field. Unlike \cite{yarom2} which studies the holographic parity-violating hydrodynamics by means of the linear response functions, we follow the standard fluid-gravity correspondence technique  \cite{Bhattacharyya:2008jc} to compute the hydrodynamic transport coefficients, including those parity-odd ones, and found some exotic properties of the latter which are yet to be understood from field theory perspective.

We employ two different parity breaking mechanisms in our system, corresponding to two different boundary conditions for the bulk pseudo scalar field. In one scenario, the pseudo scalar is sourceless on the boundary, and the parity is broken spontaneously by the nontrivial vacuum expectation value of the dual scalar operator developed below certain critical temperature $T_c$. As a result,
all parity violating coefficients are related to the order parameters for the second order phase transition of our system, and becomes non-zero below $T_c$. In this case, the boundary theory is still conformally invariant, leaving the bulk viscosity zero. Our result shows that, some of the parity-odd coefficients, such as the electric Hall conductivity ${\tilde \sigma}$ and Hall viscosity ${\widetilde \eta}_A$, go to zero with the mean field critical exponent $(T_c - T)^{1/2}$, while others, including the thermal Hall conductivity ${\widetilde \kappa}$ and the heat Hall conductivity ${\widetilde \xi}$, diverge with the critical exponent $(T_c - T)^{-3/2}$ at $T_c$. These properties are novel, and it would be interesting if they can be realized by field theory models.

In another scenario, the pseudo scalar is sourced on the boundary, and the parity of the hydrodynamic system is broken by the externally applied source of the dual boundary pseudo scalar operator. Moreover, the source also breaks the conformal symmetry. Therefore, in the hydrodynamic limit, in addition to the parity violating coefficients in the first scenario, new coefficients such as the bulk viscosity ${\zeta}$, curl viscosity ${\widetilde \zeta}_A$ and magnetic viscosity ${\widetilde \xi}_B$ also arise. We explore the holographic realization of parity breaking with both scenarios in detail, and study the finite temperature behaviors of various hydrodynamic transport coefficients of a strongly coupled system. 

Very recently, \cite{Leigh} constructs a holographic model for 2+1 dimensional fluids with background vorticity $\Omega$ and magnetic field $B$. They obtain the classical Hall conductivity inversely proportional to $B$, as expected. They also obtained the Hall viscosity inversely proportional to $\Omega$. These are different from our results, because in our model, we don't have non-vanishing external magnetic field and vorticity at the background level, i.e. in equilibrium; instead they appear as the first order perturbations. Hence, our Hall conductivity exists without the presence of the background magnetic fields, and our Hall viscosity does not depend on vorticity. Their results are obtained via the Kubo's formula, and it is interesting to realize them by the gravity-hydrodynamics correspondence.

The structure of this paper is organized as follows: in section \ref{sec:HydroRev} we briefly review the $2+1$ dimensional parity violating hydrodynamics, mostly following Ref. \cite{yarom2}. The notations subsequently used in our paper are also 
introduced here. In section \ref{sec:holographic} we present our holographic model and the ansatz for the derivative expansions of perturbations. The background solution, including two different boundary conditions for the pseudo scalar corresponding to two distinct mechanisms of breaking the boundary parity, is given in Section \ref{sec:transport}. We further reproduce and analyze the transport coefficients of parity violating hydrodynamics from the bulk gravitational theory, under the probe limit of the pseudo scalar field. Our research is concluded in Section \ref{conclusion}. The first order perturbation equations of motion of the bulk theory before taking the probe limit are listed in the Appendices.

\section{Hydrodynamics with broken parity in $2+1$ dimensions} \label{sec:HydroRev}

In the hydrodynamic limit which assumes {\it local} thermodynamical equilibrium, a system with global $U(1)$  symmetry under a non-dynamical external gauge field strength $F_{\mu\nu}^{ext}$ satisfies the following local conservation equations:
\beq
 \nabla_{\mu} T^{\mu\nu}=F^{\nu\mu}_{ext} J_{\mu}, \qquad
 \nabla_{\mu} J^{\mu}=0,
\eeq
without referring to the microscopic degrees of freedom, where $T^{\mu\nu}$ and $J^{\mu}$ are the energy-momentum tensor and the conserved $U(1)$ current respectively. 
In a 2+1 dimensional parity violating hydrodynamic system described by the velocity $u^{\mu}$, temperature $T$, and chemical potential $\mu$, the most general constitutive equations of $T^{\mu\nu}$ and $J^{\mu}$ in principle can be written down in terms of all possible combinations of these local macroscopic variables which respect the underlying symmetry and the constraint of positive entropy production \cite{yarom2}:
\begin{eqnarray}
T^{\mu \nu}
&=& 
\epsilon_0 u^{\mu} u^{\nu} + P_0 \Delta^{\mu \nu} 
-( \zeta \nabla_{\lambda} u^{\lambda} 
+ \widetilde{\zeta}_{A} \Omega + \widetilde{\zeta}_B B) \Delta^{\mu \nu}
- \eta \sigma^{\mu \nu} - \widetilde{\eta}_A \widetilde{\sigma}^{\mu \nu} ,
\label{Tmunu}
\\
J^{\mu}
&=&
\rho u^{\mu} + \sigma E^{\mu}
- \kappa \Delta^{\mu \nu} \nabla_{\nu}\frac{\mu}{T}
+ \widetilde{\sigma} \epsilon^{\mu \nu \rho} u_{\nu} E_{\rho}
+ \widetilde{\kappa} \epsilon^{\mu \nu \rho} u_{\nu} \nabla_{\rho} \frac{\mu}{T}
+ \widetilde{\xi} \epsilon^{\mu \nu \rho} u_{\nu}\nabla_{\rho} T. \label{Jmu}
\end{eqnarray}
where the combinations of the local variables are
\begin{eqnarray}
\Omega &=& -\epsilon ^{\mu \nu \rho} u_{\mu} \nabla_{\nu} u_{\rho} \qquad \mbox{(vorticity)} \nn\\
B &=& -\frac{1}{2} \epsilon^{\mu \nu \rho} u_{\mu} F_{\nu \rho}^{ext}  \qquad \mbox{(external magnetic field)}\nn\\
E^{\mu} &=& u_{\lambda} F^{\mu \lambda}_{ext} \qquad\qquad\quad \mbox{(external electric field)}\\
\Delta^{\mu \nu} &=& \eta^{\mu \nu} + u^{\mu} u^{\nu} \qquad \quad \mbox{(projector to the direction transverse to $u^\mu$)}\nn\\
\sigma^{\mu \nu} &=& \Delta^{\mu \alpha} \Delta^{\nu \beta} (\nabla_{\alpha} u_{\beta} + \nabla_{\beta} u_{\alpha} - \eta_{\alpha \beta} \nabla _{\rho} u^{\rho}) \nn\\
\widetilde{\sigma}^{\mu \nu} &=& \frac{1}{2}(\epsilon^{\mu \alpha \rho} u_{\alpha }\sigma_{\rho }{}^{\nu }+\epsilon ^{\nu \alpha \rho }u_{\alpha
}\sigma_{\rho}{}^{\mu }).\nn
\end{eqnarray}
Note that there is an ambiguity in interpreting the transport coefficients in the constitutive equations unless a particular choice of out-of-equilibrium definition of energy density, charge density, and local velocity is given, i.e. a frame is chosen. In our paper we choose the \textit{Landau frame}\footnote{Here $\{\mu,\nu\}$ run through all indices of the 2+1 dimensional spacetime, while $\{i,j\}$ only label the spatial dimensions.}, such that $u^i$ is identified to be the energy transport velocity $T^{vi}$, i.e. $T^{\mu\nu} u_{\nu} = -\epsilon_0 u^{\mu}$ and $J^{\mu} u_{\mu} = -\rho$. Then $\epsilon_0$, $\rho$, $P_0$ are interpreted as the energy density, charge density, and pressure respectively in the equilibrium configuration where $B=0$, $\Omega=0$. The local velocity is normalized as $u^{\mu} u_{\mu}=-1$. 

The coefficients in the constitutive equations are interpreted as follows. $\zeta$ and $\eta$ are the \textit{bulk} and \textit{shear viscosities}. $\sigma$ is the \textit{electric conductivity}; $\kappa = \sigma T$ is the \textit{thermal conductivity}. All these above are the dissipative transport coefficients appeared in the usual parity-even system, and they must be positive. The remaining coefficients (with tilde) arise from the parity violating effect. They are dissipationless and there is no constraint on the signs of their values. Among them, $\widetilde{\zeta}_A$ and $\widetilde{\eta}_A$ are the \textit{curl} and \textit{Hall viscosities} respectively. A cartoon depicting the physical pictures of $\widetilde{\zeta}_A$ and $\widetilde{\eta}_A$ is given in Fig. 1 of \cite{Chen}. In this paper we call $\widetilde{\zeta}_B$ the \textit{magnetic viscosity}. $\widetilde{\sigma}$ is the \textit{Hall conductivity}\footnote{We use different but equivalent parametrization in (\ref{Jmu}) compared to that in \cite{yarom2}. Our $\widetilde{\sigma}$ is in fact $\widetilde{\sigma}+\widetilde{\chi}_E$ in \cite{yarom2}.
Their $\widetilde{\chi}_E$ is equal to our $\widetilde{\sigma}-\frac{\widetilde{\kappa}}{T}$.}, which describes the Hall effect without the presence of external magnetic field\footnote{In \cite{yarom2}, $\widetilde{\sigma}$ is dubbed ``anomalous'' Hall conductivity to indicate this special feature. However, as this name clashes with the conventional notion which implies the effect is induced by anomaly, we just call it Hall conductivity to avoid confusion. Here, we remind the readers that, due to our holographic setup, all of the transport coefficients describing the response transverse to the input, including the Hall conductivity, the heat and the thermal Hall conductivity, exist without the presence of the background magnetic fields.
}. We name $\widetilde{\kappa}$ and $\widetilde{\xi}$ the \textit{thermal Hall conductivity} and 
\textit{heat Hall conductivity} respectively, which describe the transverse current due to the gradients of $\frac{\mu}{T}$ and $T$ without external magnetic field.

\cite{yarom2} refers to $\zeta$, $\eta$, $\sigma$, $\widetilde{\eta}_A$ and $\widetilde{\sigma}$ as transport coefficients, and to $\widetilde{\zeta}_A$, $\widetilde{\zeta_B}$, $\widetilde{\kappa}$ and $\widetilde{\xi}$ as thermodynamic response parameters. The difference is, in the Kubo's formula formulation, the thermodynamic response parameters are computed by setting zero-frequency before taking zero-momentum limit, while for the transport coefficients, the order is reversed. In the rest of our paper, however, we will call both of them 
``transport coefficients'' for the sake of convenience.


\section{The holographic set-up and the derivative expansion} \label{sec:holographic}

Our bulk model dual to the 2+1 dimensional parity-odd hydrodynamics 
is described by a 3+1 dimensional Einstein-Maxwell-pseudo scalar 
Lagrangian with a negative cosmological constant. The $U(1)$ gauge field 
is sourced by the gravitational background, and the neutral pseudo scalar 
couples to the gravitational and electrodynamical Chern-Simons terms\footnote{The bulk electromagnetic coupling $e^2$ is set to one.}:
\beq \label{action}
\mathcal{L} = \frac{1}{16\pi G_N}\left(R+\frac{6}{L^{2}}\right)-\frac{1}{4}F^2 + \frac{\lambda}{4}\, \theta \, \tilde{F} F
-\frac{1}{2}\,(\partial \theta)^2-V(\theta )
- \frac{\lambda'}{4} \, \theta\, \tilde{R} R .
\eeq
where
\bea
\tilde{R} R &=&  \tilde{R}^M{}_N{}^{PQ} \, R^N{}_{MPQ}\, ,  \hspace{1.5cm} \tilde{R}^M{}_N{}^{PQ} := \frac{1}{2}\, \epsilon^{PQRS} R^M{}_{NRS}, \nonumber \\
\tilde{F} F &=& \tilde{F}^{MN} F_{MN}\, , \hspace{3.1cm}
\tilde{F}^{MN} := \frac{1}{2}\, \epsilon^{MNPQ} F_{PQ}\, , \nonumber
\eea
with $\epsilon^{MNPQ}$ being the usual 4-dimensional Levi-Civita tensor. We assume that the gravitational 
and the electrodynamical Chern-Simons coupling constants are of the same order, 
in order for the parity breaking effects arising from both terms to appear at the 
same order of the probe limit we take later. For simplicity, we take $\lambda=\lambda'$ 
in (\ref{action}). 

An appropriate boundary action $S_{bdy}$ can be added to the bulk action to make the variational principle consistent on the boundary:
\beq
S_{bdy} = 2 \int d^3 x \sqrt{-h} \left( \frac{K}{16\pi G_N} -
\frac{\lambda}{2}\, \theta\, n_{A} \epsilon^{ABCD} K_{B}{}^{E}
\nabla_{C} K_{DE} \right). 
\eeq
We define $n_A$ as the outward pointing normal vector to the asymptotic boundary of AdS with unit magnitude. $K_{AB} = h_A{}^C h_B{}^D \nabla_C n_D$ is the extrinsic curvature tensor of the boundary slice, with the induced metric $h_{AB} = g_{AB}- n_A n_B$, and $K=h^{AB}K_{AB}$ is the trace of $K_{AB}$. According to the prescription of the AdS/CFT correspondence, the on-shell variation of the boundary
action with respect to the induced metric gives rise to
the energy-momentum tensor of the dual field theory. Additional boundary terms involving $\theta$ originated from $\theta \tilde{R} R$ in the bulk action in principle modifies the boundary energy-momentum tensor \cite{son2, mann}. But as long as $\theta$ is a relevant perturbation, i.e. $m^2 < 0$ such that $\theta$ is asymptotically normalizable, these additional boundary terms vanish and have no contribution to the hydrodynamic transport coefficient arising from the energy-momentum tensor. See the Appendix of \cite{son2} for detail. In this paper we focus on $m^2 < 0$, and set $16\pi G_N =1$ in the bulk action.

The 3+1 dimensional bulk coordinates are denoted by 
$x^M = (r, x^{\mu})$, where $x^{\mu}=(v,x,y)$ are the coordinates labeling the boundary slice.

The parity in the bulk Lagrangian (\ref{action}) is conserved. 
Parity violation in the boundary theory can arise either spontaneously from the $\theta$ condensate corresponding to the bulk solutions which satisfy the sourceless boundary condition, or ``explicitly'' from the bulk pseudo scalar with a source on the boundary.  The potential $V(\theta)$ in our model takes the form\footnote{It is known that with the potential (\ref{V}), a neutral hairy scalar can only satisfy the sourceless boundary condition in a chargeless AdS-Schwarzschild background for certain range of $c$ in which the positive energy theorem is violated \cite{Hertog:2006rr}. Introduction of the gauge fields sourced by the gravitational background eliminates this problem.

In this paper, for the case where $\theta$ is asymptotically sourceless, we focus on the regime $-\frac{9}{4} < m^2 < -\frac{3}{2}$, as in order for $\theta$ to have a non-vanishing vev on the boundary, it has to violate the Breitenlohner-Freedman bound in the near horizon region, which is $m^2=-\frac{3}{2}$ for our black brane background whose near horizon geometry at $T=0$ is $AdS_2$. For the case where $\theta$ is sourced on the boundary, we take the $m^2$ range such that $\theta$ is dual to the relevant deformation in the boundary theory, i.e. $-\frac{9}{4} < m^2 < 0$.}
\beq \label{V}
V(\theta)=\frac{1}{2} m^2 \theta^2+ \frac{1}{4}\, c\: \theta^4.
\eeq 
The equations of motion arising from (\ref{action}) and (\ref{V}) are
\bea 
R_{MN}-\frac{1}{2} g_{MN} R + \Lambda g_{MN} - \lambda C_{MN} &=& T_{MN} ({\theta})
 + T_{MN} (A),  \nonumber \\
\nabla_M F^{MN} &=& \lambda\: \partial_M \theta\, \tilde{F}^{MN}, \label{EOM} \\ 
\nabla^2 \theta &=& \frac{dV}{d \theta} + \frac{\lambda}{4}\, \tilde{R} R -
\frac{\lambda}{4}\, \tilde{F} F, \nonumber
\eea
where
\bea
T_{MN}(\theta) &=& \frac{1}{2} \, \partial_M \theta \, \partial_N \theta -
\frac{1}{4}\, g_{MN} (\partial \theta)^2 - \frac{1}{2} \, g_{MN} V(\theta ),  \nonumber \\
T_{MN}(A) &=& \frac{1}{2}\, F_M{}^A F_{NA} - \frac{1}{8}\, g_{MN} F_{AB}F^{AB},  \nonumber \\
C^{MN} &=&   \epsilon^{PQS(M} \nabla_S R^{N)}{}_Q\,\nabla_P \theta
  + \tilde{R}^{Q(MN)P} \nabla_P \nabla_Q \theta . \nonumber
\eea
$C_{MN}$ is a symmetric traceless tensor analogous to the Cotton tensor in three dimensions, and the symmetrization is defined as $\tilde{R}^{Q(MN)P}:=\frac{1}{2}(\tilde{R}^{QMNP}+\tilde{R}^{QNMP})$.

The background solution to the bulk equations of motion (\ref{EOM}), which describes a uniform ideal fluid on the boundary with velocity $u^{\mu}$, temperature $T$ and chemical potential $\mu$, is given by the following boosted charged brane ansatz
\bea
ds^2&=&-2\,H(r,M,Q)\, u_{\mu} dx^{\mu} dr
  - r^{2} f(r,M,Q)\, u_{\mu} u_{\nu}dx^{\mu} dx^{\nu}
  + r^{2} \Delta_{\mu\nu} dx^{\mu} dx^{\nu} \: := \:ds^2{}^{\:(0)} , \nn \\
\theta &=& \theta (r,M,Q)\: := \:\theta^{(0)},  \label{BG} \\
A &=& \left[A(r,M,Q)\, u_{\mu} + A^{ext}_{\mu}\right]dx^{\mu} \: := \: A^{(0)} , \nn \\
A^{ext}_{\mu} &=& (A_v^{ext},A_x^{ext}, A_y^{ext})
, \nn
\eea
where $u^{\mu}$ is the three-velocity on the boundary with $u_{\mu} u^{\mu} = -1$, $\mu$ is given by the difference between the boundary and the horizon value of $A_v$, $T$ is the black hole temperature, and $\Delta_{\mu\nu}=\eta_{\mu\nu} + u_{\mu} u_{\nu}$ is the projector to the directions transverse to $u^{\mu}$. Here we also turn on the constant external gauge fields $A^{ext}_{\mu}$, besides the dynamical ones. Since $A^{ext}_{\mu}$ are constants in our setup, they have no contribution to the background bulk dynamics. However, the external electric and magnetic fields will arise from perturbing $A^{ext}_{\mu}$, which will be clear shortly. The superscript $(0)$ denotes the background ansatz with respect to the perturbations and corrections introduced later. (See (\ref{fullans}).) We also choose the Landau frame for our bulk background.

For 2+1 dimensional point of view, by promoting the uniform local variables $u^{\mu}$, $T$ and $\mu$ to slow-varying functions of spacetime without breaking local conservation laws, one can add dissipative and non-dissipative correction terms with various transport coefficients to the energy-momentum tensor and the current of the uniform ideal hydrodynamics. One can also turn on external electric and magnetic fields which also contribute to the corrections to $T^{\mu\nu}$ and $J^{\mu}$. 

In terms of the dual gravity, this is equivalent to promoting the constant bulk quantities $M, Q, u^{\mu}$ and $A^{ext}$ to slow-varying functions of the boundary coordinates $x^{\mu}$, and then determining their effective dynamics. Due to this field promotion, the original bulk fields undergo derivative expansions, which together with the correction ansatz solve the equations of motion order by order. By solving all the equations of motion at each order, the boundary hydrodynamics in terms of these slow-varying fields are realized by the standard AdS/CFT correspondence, and the transport coefficients are extracted from the asymptotic behaviors of the bulk field solutions\cite{Bhattacharyya:2008jc}. For the charged fluid case, the holographic description is summarized in \cite{Banerjee:2008th}. We will apply this approach to our bulk theory (\ref{action}) to study the dual parity violating hydrodynamics on the boundary. 

In the co-moving frame, at the origin of the boundary coordinates $x^{\mu}=0$  we have $u^{\mu}=(1,0,0)$. The promoted slow varying fields $u^{\mu}(x^{\nu}), M(x^{\nu}), Q(x^{\nu}), A_{\mu}^{ext}(x^{\nu})$ are expressed in terms of the Taylor expansions with respect to the origin:
\bea
u^{\mu}(x^{\nu}) &=& (1,\;x^{\nu} \partial_{\nu} \beta^i) = (1, \;\delta \beta^i),  \nn \\
M(x^{\nu}) &=& M + x^{\nu}\partial_{\nu}M, \label{DexpC} \\
Q (x^{\nu}) &=& Q + x^{\nu} \partial_{\nu } Q, \nn \\
A^{ext}_{\mu}(x^{\nu}) &=& A^{ext}_{\mu} + x^{\nu} \partial_{\nu} A^{ext}_{\mu}. \nn
\eea
As a result, the bulk metric components $f(r,M,Q), H(r,M,Q)$, dynamical gauge fields $A (r,M,Q)$, and pseudo scalar $\theta (r,M,Q)$ also receive derivative expansions,
\beq \label{Dexp}
 \delta F(r,M,Q) = \frac{\partial F}{\partial M}\, x^{\mu} \partial_{\mu} M
   + \frac{\partial F}{\partial Q}\, x^{\mu} \partial_{\mu} Q, 
\eeq
where $F = \{f, H, A, \theta\}$. Since such perturbed ansatz is no longer a solution to the original equations of motion, the following corrections to the bulk fields must be introduced to compensate the deviation from the original solution due to derivative expansions (\ref{DexpC}) and (\ref{Dexp})
:
\bea
ds^2{}^{\:(1)} &=& r^2 k(r)\, dv^2 + 2 H h(r)\, dv dr + 2 r^2 j_i(r)\, dv dx^i
   - r^2 h(r)\, dx^i dx^i + r^2 \alpha_{ij}(r)\, dx^i dx^j\, ,  \nonumber \\
A^{(1)} &=& a_v(r)\,dv + a_i(r)\,dx^i \, , \label{corr} \\
\theta^{(1)} &=& \varphi (r)\, , \nonumber   
\eea
where $\alpha_{ij}$ is symmetric traceless.

Combining (\ref{BG})$\sim$(\ref{corr}), the overall charged black brane ansatz up to first order perturbation reads
\bea  
ds^2 &=& ds^2{}^{\:(0)} \nn \\
     && +\, \epsilon \left[ - r^2 \delta f\: dv^2 + 2\, \delta H\: dvdr
          - 2\, r^2 (1-f(r))\, \delta \beta^i\: dv dx^i
          - 2\,H(r)\, \delta \beta^i\: dr dx^i \right] \nn\\
     && +\, \epsilon\; ds^2{}^{\:(1)} ,  \label{fullans} \\
A &=& A^{(0)} + \epsilon \,( - \delta A\: dv 
      + A(r)\, \delta \beta^i \: dx^i + A^{(1)}), \nn \\
\theta &=& \theta^{(0)} + \epsilon\, (\delta \theta + \theta^{(1)}).  \nn
\eea
The parameter $\epsilon$ is introduced to keep track of the orders of derivatives and corrections. The superscripts $(0)$ and $(1)$ in equations (\ref{BG}) and (\ref{corr}) label the $O(\epsilon^0)$ and $O(\epsilon^1)$ ansatz respectively, and the whole ansatz (\ref{fullans}) solve the bulk equations of motion up to $O(\epsilon^1)$. 
The same procedure can be generalized straightforwardly to the higher order calculation.

In the next section, we will present the solutions at the probe limit of the pseudo scalar field, and extract the corresponding transport coefficients for the boundary fluid.


\section{Transport coefficients of parity violating fluids in $2+1$ dimensions from gravity} \label{sec:transport}

In this section, we will solve the bulk equations of motion up to $O(\epsilon)$ perturbations, and extract the first order transport coefficients of the dual hydrodynamics. 

It is clear from (\ref{EOM}) that the gravitational background receives backreaction from nontrivial pseudo scalar profile. Despite that the full analytic solution with backreaction is not yet known, we can still explore the dual hydrodynamics in the probe limit of the pseudo scalar field
\beq \label{probe}
\theta \to \lambda \theta\, ,\qquad V(\theta) \to \lambda V(\theta)\, , \qquad \lambda \to 0 \, .
\eeq
We will see that, in vector and tensor modes, by taking the probe limit as in (\ref{probe}), we are able to separate parity-even transport coefficients which can be see at $O(\lambda^0)$, from the parity-odd transport ones which only appear at higher order in $\lambda$.

We first present the background solution in the probe limit. In the Einstein and Maxwell equations, since the leading order term are of $O(\lambda^0)$ while the $\theta$ terms contribute at $O(\lambda^2)$, the pseudo scalar decouples at the leading order of the $\lambda$ expansion. The background is therefore a charged black brane in $AdS_4$,
\bea
ds^2{}^{\:(0,\tilde{0})} &=& 2\,dvdr-r^{2}f(r)dv^2+r^2(dx^2+dy^2), \nn \\
f(r)&=& 1-\frac{M}{r^3}+\frac{Q^2}{r^{4}}, \qquad H(r) = 1, \label{blackbrane}\\ 
A^{(0,\tilde{0})} &=& -\frac{2 Q}{r}dv, \nn
\eea
where another superscript $(\tilde{n})$ is employed to label the order of $\lambda$, while the untilded one $(n)$ denote the order of $\epsilon$.  The horizon is at $r=r_H$, and\footnote{
The action and equations of motion have the following scaling symmetry: 
\bea
r \!&\to&\! b\,r,\quad (v,x,y)\to \frac{1}{b}\,(v,x,y),  \nn \\
Q \!&\to&\! b^{2}Q, \hspace{1.05cm} M \to b^{3}M,  \nn \\
\theta \!\!&\to&\!\! \theta, \qquad \qquad\: A \to A, \qquad \qquad f \to f.  \nn
\eea
One can make use of this invariance to rescale the horizon to $r_H =1$ by setting $b=1/r_H,\, M=1+3\kappa$ and $Q^2=3\kappa$, but the independence between temperature $T$ and the chemical potential $\mu$ (or the charge $Q$) in the boundary theory will be lost.} $M$, $Q$ are the black brane mass and electric charge respectively. The metric is asymptotically $AdS_4$ with the $AdS$ radius taken to be 1. The temperature and the chemical potential are
\beq \label{T}
T = \frac{3r_H}{4\pi}\left(1-\frac{Q^2}{3 r_H^4}\right),\qquad \mu = \frac{2Q}{r_H}.
\eeq

The probe limit (\ref{probe}) implies that the scalar field is in fact $\theta = \theta^{(0,\tilde{1})}$. Since (\ref{blackbrane}) gives rise to vanishing $\tilde{R}R$ and $\tilde{F}F$ terms, the Klein-Gordon equation in (\ref{EOM}) reduces to 
\beq \label{KGb}
\theta''+(\frac{f'}{f}+\frac{4}{r})\,\theta'-\frac{dV/d\theta}{r^{2}f}=0.
\eeq
Since $f(r=r_H) = 0$, it follows straightforwardly from the above equation that on the horizon $\theta$ is subject to the condition
\beq \label{thetareg}
 \theta\,{}' (r=r_H) = \left. \frac{dV/d\theta}{r^2 f'}\right|_{r=r_H}.
\eeq
On the other hand, the pseudo scalar behaves asymptotically as
\beq \label{thetaasym}
\theta = \frac{J^{(0)}}{r^{\Delta_-}}+\frac{\left\langle \mathcal{O}\right\rangle^{(0)}}{r^{\Delta_+}}+\cdots,
\qquad\quad \Delta_{\pm} = \frac{3}{2} \pm \sqrt{\frac{9}{4}+m^2}. 
\eeq
For $- \frac{5}{4} < m^2 < 0$, $J^{(0)}$ is identified as the boundary source of $\theta$, while $\left\langle \mathcal{O}\right\rangle^{(0)}$ is the response (or, condensate) on the boundary dual to the source $J^{(0)}$. For $-9/4 < m^2 < -5/4$, one is free to choose either $J^{(0)}$ or $\left\langle \mathcal{O}\right\rangle^{(0)}$ as the source and the other as condensate \cite{Klebanov:1999tb}. In this paper we choose $J^{(0)}$ as the source and $\left\langle \mathcal{O}\right\rangle^{(0)}$ as the condensate.

In our model, we consider two types of asymptotic boundary conditions for $\theta$, corresponding to two mechanisms to break the parity in the boundary theory:

\noindent{\it (i) Boundary condition with a source:} 

One can turn on the boundary source $J^{(0)}$ for $\theta$, such that the condensate $\left\langle \mathcal{O}\right\rangle^{(0)}$ is determined by the regularity condition on the horizon. In the analysis throughout this paper, We set $J^{(0)}=1$ for convenience. In this case, the boundary parity and conformal symmetry are broken by the source $J^{(0)}$. 

\noindent{\it (ii) Sourceless boundary condition:} 

\begin{figure}[tbp]
\begin{center}
\includegraphics[width=0.45\textwidth]{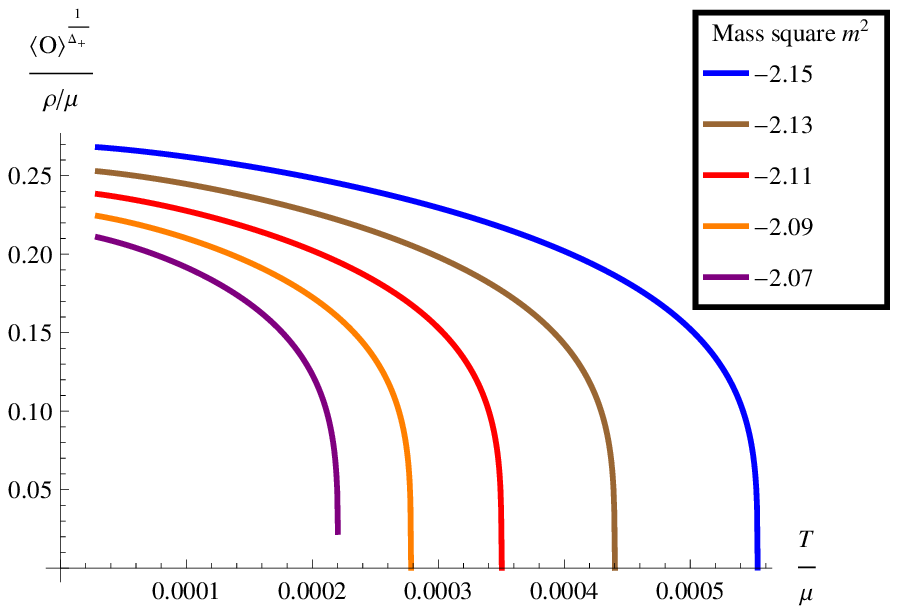}
\includegraphics[width=0.49\textwidth]{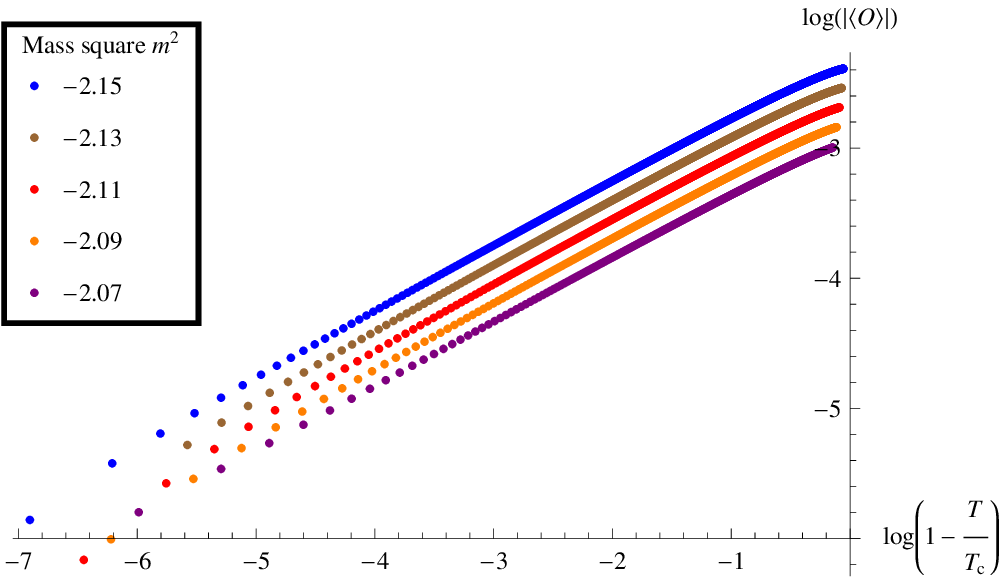}
\caption{$\frac{(\left\langle \mathcal{O}\right\rangle^{(0)})^{1/\Delta_+}}{\rho/\mu}$ as a function of $T/\mu$ with linear scale (on the left) and log scale (on the right) for various $m^2$. In these figures $\theta$ is asymptotically sourceless, i.e. $J^{(0)}=0$ in (\ref{thetaasym}). $\frac{(\left\langle \mathcal{O}\right\rangle^{(0)})^{1/\Delta_+}}{\rho/\mu}$ vanishes at and above the critical temperature, with a critical exponent equal to 0.5, as being indicated by the right plot.} \label{fig:O}
\end{center}
\end{figure}

The other scenario is, the source $J^{(0)}$ can be switched off consistently, such that the parity and conformal symmetry in the boundary theory are broken spontaneously by the nontrivial vev of $\left\langle \mathcal{O}\right\rangle^{(0)}$ developed below certain critical temperature depending on $m^2$. In this case, from the Klein-Gordon equation (\ref{KGb}), one can derive the mean field description of $\left\langle \mathcal{O}\right\rangle^{(0)} \sim (Tc-T)^{0.5}$ near $T_c$. This is confirmed by the numerical results\footnote{We take $c=0.5$ in the potential $V(\theta)$ throughout the numerical analysis in this paper.} in Fig. \ref{fig:O}. 

Having the unperturbed background solution in hand, in the following we solve the $O(\epsilon)$ equations of motion, in terms of the scalar, vector, tensor modes with respect to the boundary spatial $SO(2)$ symmetry in the probe limit. The full $O(\epsilon)$ equations with backreaction are presented in the Appendices for interested readers. We will denote the $O(\epsilon)$ Einstein, Maxwell and Klein-Gordon equations by $E_{MN}^{(1)}$, $M_N^{(1)}$, and $KG^{(1)}$ respectively.


\subsection{Scalar mode} \label{scalarmode}

In the probe limit (\ref{probe}) and $\varphi \to \lambda \varphi$, there are four dynamical equations for the scalar modes arising from $E_{rr}^{(1)}, H(r) M^v{}^{(1)}, \frac{r^2 f}{H} E_{rr}^{(1)} + E_{vr}^{(1)}$ and $KG^{(1)}$. The first three equations are
\bea
\frac{1}{r^4} (r^4 h')' &=& O(\lambda^2) , \nn \\
\frac{1}{r^2} (r^2 a_v{}')' + 2 A' h' &=& O(\lambda^2), \\
\frac{1}{r^2} (r^3 k)' + \frac{1}{2r^2}\Big( (r^4 f)' h \Big)'
-\frac{1}{2} A' a_v{}' &=& \frac{2}{r}\, \partial_i \beta_i +O(\lambda^2). \nn 
\eea
Substituting $f(r)$ and $A(r)$ from $O(\epsilon^0)$ solution (\ref{blackbrane}) into the equations above, one obtains
\bea
h(r) &=& c_1 - \frac{c_2}{3r^3} + O(\lambda^2), \nn \\
a_v (r) &=& c_3 - \frac{c_4}{r} - \frac{c_2 Q}{3 r^4}+ O(\lambda^2), 
\label{hak}\\
k(r) &=& -2c_1 + \frac{c_5}{r^3} + \frac{c_4 Q}{r^4} 
- \frac{c_2 M}{6 r^6} +\frac{c_2 Q^2}{3 r^7}
   + \frac{1}{r} \,\partial_i \beta_i  +O(\lambda^2), \nn
\eea
where $c_1, \ldots , c_5$ are arbitrary constants. 

In our model, the integration constants $c_1, c_3$ are set to zero by choosing the normalizable boundary conditions for $h(r)$ and $a_v (r)$. $c_4, c_5$ takes zero value due to the fact we choose the Landau frame such that $\pi^{\mu \lambda} u_{\lambda}=0$ and $\nu^{\mu} u_{\mu}=0$, where $\pi^{\mu \lambda},\; \nu^{\mu}$ are $O(\epsilon)$ corrections to the energy momentum tensor $T^{\mu \lambda}$ and the current $j^{\mu}$ respectively\cite{yarom2}. 
$c_2$ can be absorbed by redefinition of the coordinate $r$ and set to zero \cite{Bhattacharyya:2008jc, Banerjee:2008th}. Hence,
\bea \label{havk}
h(r) = 0, \qquad a_v (r) = 0, \qquad k(r) = \frac{1}{r}\, \partial_i \beta_i .
\eea
With (\ref{havk}), the constraint equations (\ref{smodecons1}) and (\ref{smodecons2}) arising from $g^{rr} E_{vr}^{(1)} + g^{vr} E_{vv}^{(1)}$ and $M^r{}^{(1)}$ reduce to
\bea
\partial_v Q &=& - Q\, \partial_i \beta_i + O(\lambda^2), \label{scalarconsprobe} \\
\partial_v M &=& -\frac{3 M}{2}\, \partial_i \beta_i + O(\lambda^2). \nn
\eea
By substituting in (\ref{havk}) and (\ref{scalarconsprobe}), the fourth dynamical equation, i.e. $KG^{(1)}$ of $O(\lambda)$, reads
\bea
\frac{1}{r^2} \big(r^4 f \varphi' \big)' 
- \frac{d^2 V}{d \theta^2} \varphi 
&=&
\left[
\frac{1}{r}\,(r^2 \theta')'+\frac{3M}{r}\left(r \frac{\partial \theta}{\partial M}\right)' + \frac{2Q}{r}\left( r \frac{\partial \theta }{\partial Q}\right)'
\right] \partial_i \beta_i \nn
\\
&&+
\left[
\frac{1}{4r^2} \left(r^4 f'^2\right)' - \frac{2 A A'}{r^2}
\right] (\partial_x \beta_y - \partial_y \beta_x)
-\frac{2A'}{r^2} B , \label{KG1}
\eea
where $B = F_{xy}^{ext} =\partial_x A_y^{ext} - \partial_y A_x^{ext}$. Note that unlike (\ref{hak}), the $KG^{(1)}$ equation (\ref{KG1}) describes parity violating effects.

To solve $KG^{(1)}$, we decompose $\varphi$ into the ``bulk'', ``curl'', and `magnetic'' parts which are proportional to $\partial_i \beta_i$, $\epsilon_{ij}\, \partial_i \beta_j$, $B$ respectively, as they are independent from each other:
\beq \label{phidecomp}
 \varphi (r) = \varphi_b (r)\, \partial_i \beta_i
 + \varphi_c (r) \, (\partial_x \beta_y - \partial_y \beta_x) 
 + \varphi_{B} (r) \, B. 
\eeq
such that
\bea
\frac{1}{r^2} \big(r^4 f \varphi_b' \big)' 
- \frac{d^2 V}{d \theta^2} \varphi_b &=&
\frac{1}{r}\,(r^2 \theta')'+\frac{3M}{r}\left(r \frac{\partial \theta}{\partial M}\right)' + \frac{2Q}{r}\left( r \frac{\partial \theta }{\partial Q}\right)' ,\nn \\
\frac{1}{r^2} \big(r^4 f \varphi_c' \big)' 
- \frac{d^2 V}{d \theta^2} \varphi_c 
&=& \frac{1}{4r^2} \left(r^4 f'^2\right)' - \frac{2 A A'}{r^2}, \\
\frac{1}{r^2} \big(r^4 f \varphi_{B}' \big)' 
- \frac{d^2 V}{d \theta^2} \varphi_{B} 
&=& -\frac{2 A'}{r^2}. \nn
\eea
$\varphi_b$, $\varphi_c$ and $\varphi_B$ are subject to the regularity condition on the horizon,
\bea
\varphi_b{}' (r_H) &=& \frac{1}{r^2 f'} \Bigg[
\frac{1}{r} \frac{d}{dr} (r^2 \theta') 
+\frac{3M}{r}\frac{d}{dr}\left(r\frac{\partial \theta}{\partial M}\right)
+\frac{2Q}{r}\frac{d}{dr}\left(r\frac{\partial \theta}{\partial Q}\right)
+\frac{d^2 V}{d \theta^2} \varphi_b
 \Bigg]_{r=r_H}, \nn \\
\varphi_c{}' (r_H) &=& \frac{1}{r^2 f'} \Bigg[ 
\frac{1}{4r^2}\frac{d}{dr}\left(r f'^2\right)
-\frac{2AA'}{r^2}
+\frac{d^2 V}{d \theta^2} \varphi_c
\Bigg]_{r=r_H}, \\
\varphi_B{}' (r_H) &=& \frac{1}{r^2 f'} \Bigg[ 
-\frac{2A'}{r^2}
+\frac{d^2 V}{d \theta^2} \varphi_B
\Bigg]_{r=r_H}, \nn 
\eea
and asymptotically it behaves as
\bea
\theta &=& \theta^{(0,\widetilde{1})} + \epsilon\, \theta^{(1,\widetilde{1})}
\: = \: \frac{J}{r^{\Delta_-}}+\frac{\left\langle \mathcal{O}\right\rangle}{r^{\Delta_+}} \nn \\
&=&
\frac{J^{(0)}+\epsilon (J_b^{(1)} \partial_i \beta_i + J_c^{(1)} \epsilon_{ij}\, \partial_i \beta_j + J_B^{(1)} B)}{r^{\Delta_-}} \nn \\
&&\quad
+\frac{\left\langle \mathcal{O}\right\rangle^{(0)} + \epsilon (\left\langle \mathcal{O}\right\rangle_b^{(1)} \partial_i \beta_1 + \left\langle \mathcal{O}\right\rangle_c^{(1)} \epsilon_{ij} \, \partial_i \beta_j + \left\langle \mathcal{O}\right\rangle_B^{(1)} B)}{r^{\Delta_+}} \nn\\
&=& \frac{1}{r^{\Delta_-}} + 
\frac{\left\langle \mathcal{O}\right\rangle^{(0)} + \epsilon (\left\langle \mathcal{O}\right\rangle_b^{(1)} \partial_i \beta_1 + \left\langle \mathcal{O}\right\rangle_c^{(1)} \epsilon_{ij}, \partial_i \beta_j + \left\langle \mathcal{O}\right\rangle_B^{(1)} B)}{r^{\Delta_+}} , \label{thetavisc}
\eea
where $\theta^{(1,\widetilde{1})} = \varphi$, and we have chosen the boundary condition $\{J^{(0)}=1,\ J^{(1)}=0\}$ such that $\{\left\langle \mathcal{O} \right\rangle^{(0)},\ \left\langle \mathcal{O} \right\rangle^{(1)}\}$ are determined by the regularity condition of $\theta$ and $\varphi$ on the horizon respectively. The boundary conformal symmetry and the parity are broken by the source $J^{(0)}$. We choose $J^{(1)}=0$, because the source on the the boundary is specified by $O(\epsilon^0)$ background and we don't want to be modified by the bulk perturbation\footnote{On the other hand, if one chooses both $J^{(0)}=0$ and $J^{(1)}=0$, the viscosities arising from the scalar mode are just trivially zero in (\ref{traceT}).}. Then, the bulk viscosity $\zeta$, curl viscosity $\widetilde{\zeta}_A$, and magnetic viscosity $\widetilde{\zeta}_B$ can be extracted from the following expression of the trace of the boundary energy-momentum tensor \cite{Bianchi:2001, Yarom:2009}:
\begin{figure}[tbp]
\begin{center}
\includegraphics[width=0.49\textwidth]{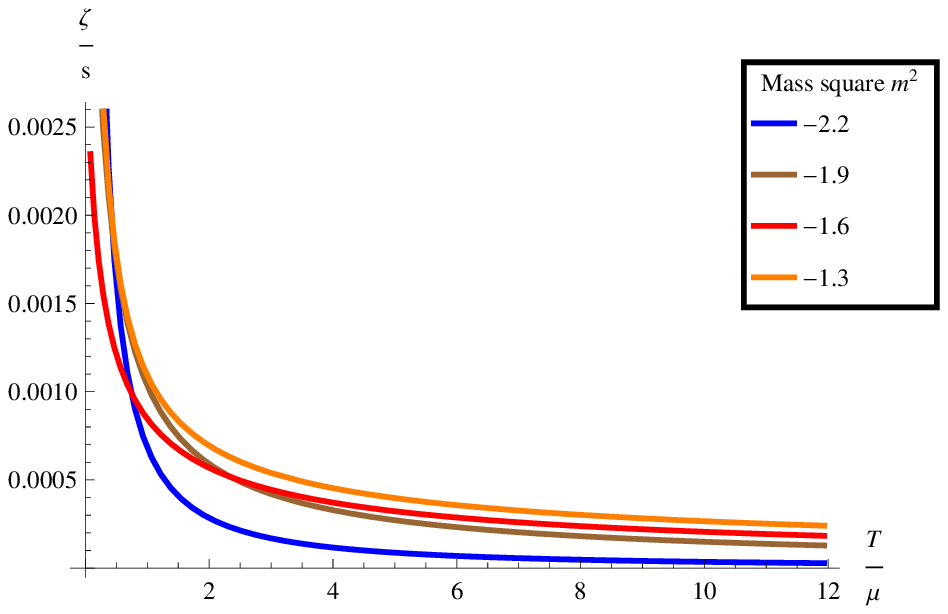}
\includegraphics[width=0.49\textwidth]{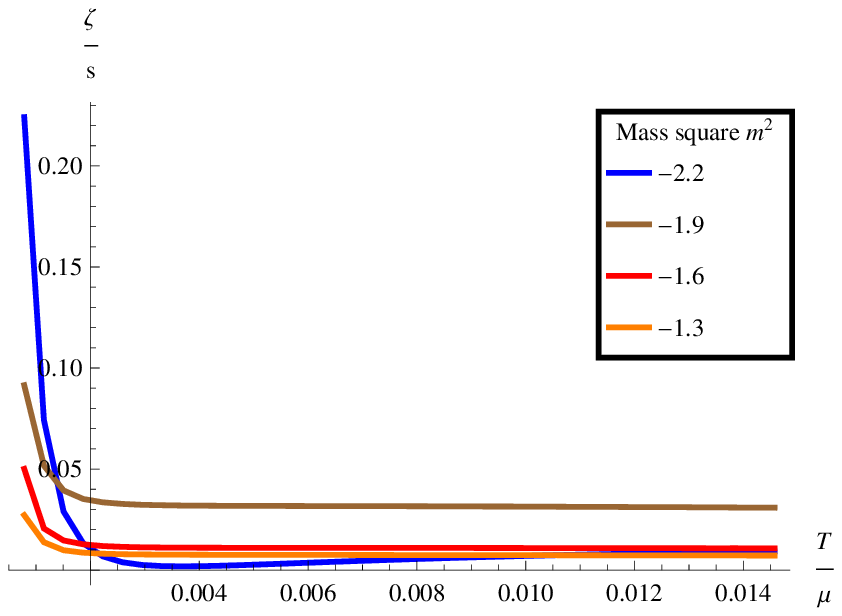}
\caption{$\zeta/s$ versus $T/\mu$ for various $m^2$. The right figure is for the very small $T/\mu$. The boundary conditions of $\theta$ and $\varphi$ underlying these numerical results are summarized below equation (\ref{thetavisc}).} \label{fig:zeta}
\includegraphics[width=0.49\textwidth]{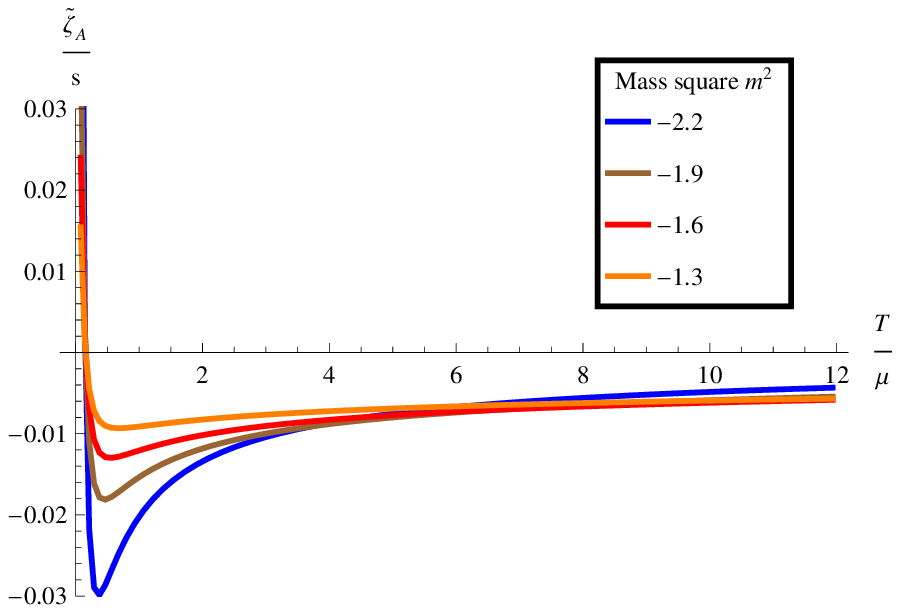}
\includegraphics[width=0.49\textwidth]{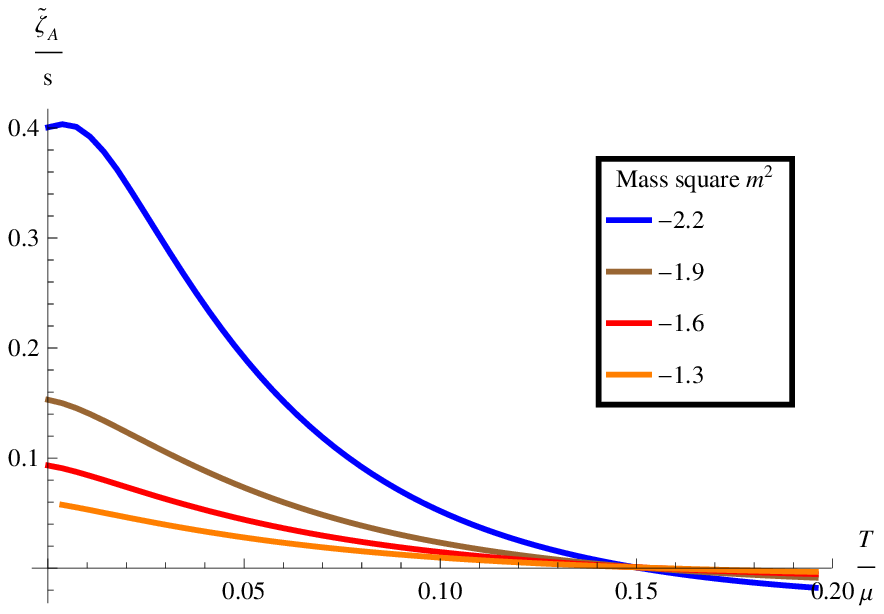}\caption{$\widetilde{\zeta}_A/s$ versus $T/\mu$ for various $m^2$. The right figure is for the $T/\mu$ range close to zero. The boundary conditions of $\theta$ and $\varphi$ underlying these numerical results are summarized below equation (\ref{thetavisc}).} \label{fig:zetaA}
\includegraphics[width=0.49\textwidth]{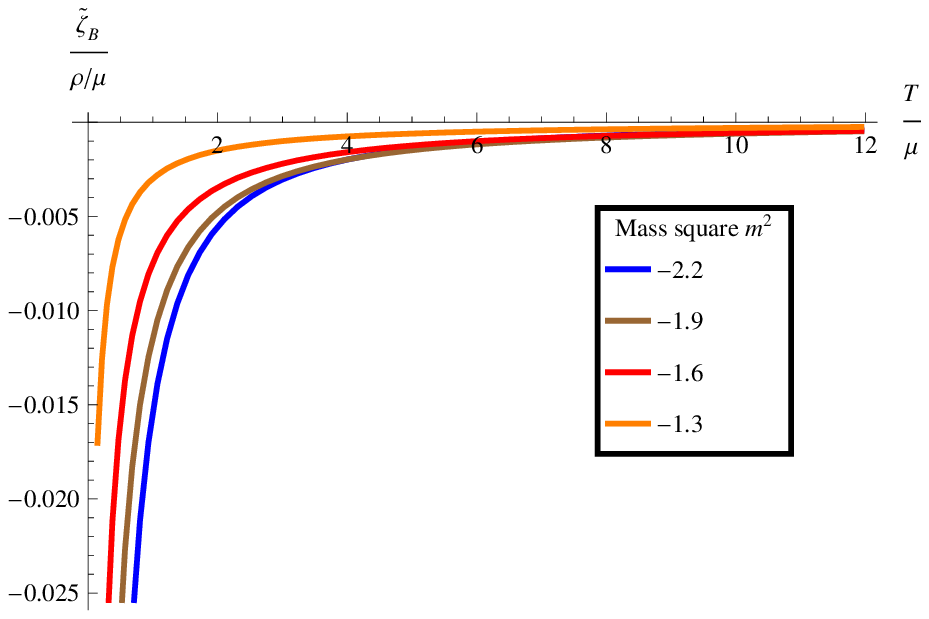}
\includegraphics[width=0.49\textwidth]{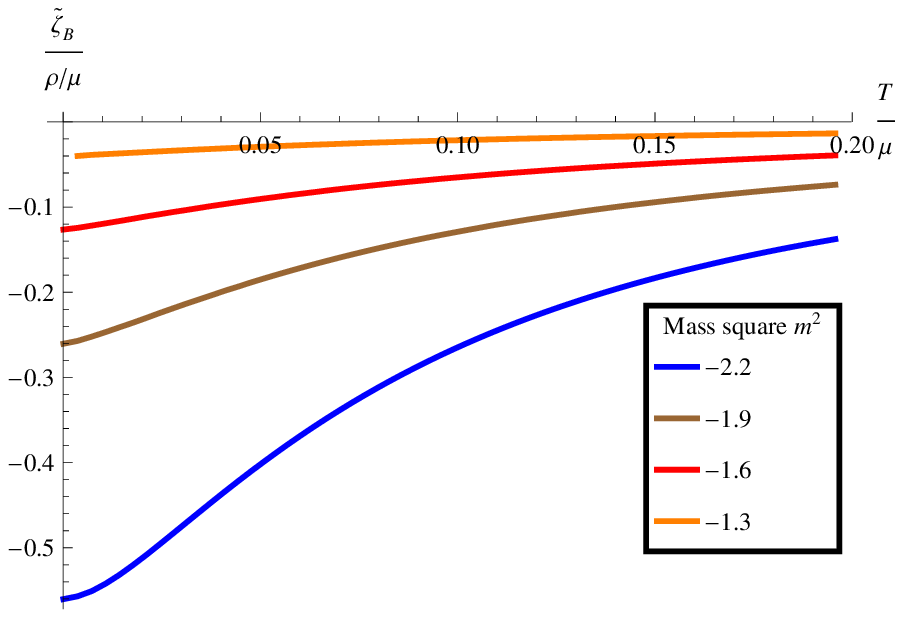}\caption{$\frac{\widetilde{\zeta}_B}{\rho/\mu}$ versus $T/\mu$ for various $m^2$. The right figure is for very small $T/\mu$. The boundary conditions of $\theta$ and $\varphi$ underlying these numerical results are summarized below equation (\ref{thetavisc}).} \label{fig:zetaB}
\end{center}
\end{figure}
\bea
\left\langle T^{\mu}{}_{\mu}\right\rangle^{(1)} 
&=&  
\Delta_- \big(J^{(0)}\left\langle \mathcal{O}\right\rangle^{(1)} +  J^{(1)}\left\langle \mathcal{O}\right\rangle^{(0)}\big) \nn\\
&=&
\Delta_- J^{(0)}
\Big( 
\left\langle \mathcal{O}\right\rangle_b^{(1)} 
\partial_i \beta_i + \left\langle \mathcal{O}\right\rangle_c^{(1)} \epsilon_{ij}\, \partial_i \beta_j + \left\langle \mathcal{O}\right\rangle_B^{(1)} B 
\Big)\label{traceT}\\
&=&
 2\, \zeta \,\partial_i \beta_i + 2\, \widetilde{\zeta}_A\, \epsilon_{ij}\, \partial_i \beta_i + 2\, \widetilde{\zeta}_B B, \nn
\eea
where 
\bea 
\zeta &=& \frac{\Delta_-}{2} 
\left\langle \mathcal{O}\right\rangle_b^{(1)}, \qquad \mbox{(bulk viscosity)}\nn \\
\widetilde{\zeta}_A &=& \frac{\Delta_-}{2}  
\left\langle \mathcal{O}\right\rangle_c^{(1)}, \qquad \mbox{(curl viscosity)} \label{zeta} \\
\widetilde{\zeta}_B &=& \frac{\Delta_-}{2}  
\left\langle \mathcal{O}\right\rangle_B^{(1)}. \qquad \mbox{(magnetic viscosity)}\nn
\eea

The numerical results for the dimensionless combinations $\frac{\zeta}{s}$, $\frac{\widetilde{\zeta}_A}{s}$ and $\frac{\widetilde{\zeta}_B}{\rho/\mu}$ are in Fig. \ref{fig:zeta}$\sim$\ref{fig:zetaB}. Our numerical analysis shows that $\zeta/s$ is positive, and diverges at $T/\mu=0$. $\widetilde{\zeta}_A/s$ is negative, and becomes more negative toward smaller $T/\mu$ before it starts to increase and eventually flips sign at very small $T/\mu$, tending to certain finite value at $T/\mu=0$. The point at which $\widetilde{\zeta}_A/s$ crosses the $T/\mu$ axis in Fig. \ref{fig:zetaA} turns out independent of  $m^2$. This property is novel and the underlying physics remains unclear. However, since the curl viscosity is dissipationless, the change of sign is allowed, as there is no constraint on the sign of $\widetilde{\zeta}_A$. 
On the other hand, $\frac{\widetilde{\zeta}_B}{\rho/\mu}$ is negative and decreases monotonically as $T/\mu$ reduces, until it hits certain finite value at $T/\mu=0$. Note that all of $\frac{\zeta}{s}$, $\frac{\widetilde{\zeta}_A}{s}$ and $\frac{\widetilde{\zeta}_B}{\rho/\mu}$ vanish as $T/\mu \to \infty$. This is expected because they are related to the asymptotic behavior of $\theta$. 
As the temperature is much higher than the scales set by the source $J^{(0)}$ and the chemical potential $\mu$, i.e. $T>>J^{(0)}$ and $T>>\mu$, $J^{(0)}$ and $\left\langle \mathcal{O}\right\rangle^{(1)}$ become negligible compared to $T$, and therefore these three transport coefficients vanish.  


\subsection{Vector mode} \label{vectormode}


\subsubsection {Parity-even part}

For the $O(\epsilon)$ vector mode perturbations, there are two dynamical equations from $E^{ri}{}^{(1)}$ and $r^2 M^i{}^{(1)}$. In the probe limit they reduce to
\bea
\frac{1}{2r^2} \frac{d}{dr}\big( r^4 j_i'\big) 
- \frac{1}{2} A' a_i' &=& -\frac{\partial_v \beta_i}{r} + O(\lambda^2), \label{vmodeeom1} \\
\frac{d}{dr}\big(r^2 f a_i' \big) - r^2 A' j_i' 
&=& - \frac{\partial A'}{\partial Q}\partial_i Q - A'\partial_v \beta_i + O(\lambda^2), \label{vmodeeom2}
\eea
where $r^2 j_i = g_{vi}^{(1)}$. The dynamical equations (\ref{vmodeeom1}) and (\ref{vmodeeom2}) describe a parity-even system. In the probe limit, the constraint equation from $\frac{r^2 f}{H} E_{ri}^{(1)} + E_{vi}^{(1)}$ reads
\beq \label{vconsprobe}
\partial_i M = -3\, M \, \partial_v \beta_i + 2\, Q\, E_i + O(\lambda^2),
\eeq
where $E^i = F_{vi}^{ext}$ is the external electric field.

$a_i$ and $j_i$ can be solved analytically from (\ref{vmodeeom1}) and (\ref{vmodeeom2}). After integrating both equations once, one can write down the decoupled differential equation for $j_i$:
\beq \label{jeq}
(r^6 f) j_i''+(4 r^5 f) j_i' - 4 Q^2 j_i 
= (-2 r^3 f + \frac{4Q^2}{r})\partial_v \beta_i 
+\frac{4Q}{r} \partial_i Q + 2 Q b ,
\eeq
where $b$ is an undetermined integral constant. Observing that the left hand side can be written into a total derivative $r^2 (r^4 f^2 (j_i/f)')'$, one obtains the solution for $j_i$,
\beq \label{jgeneral}
 j_i = -f \int_{\infty}^r ds \left\{ \frac{1}{s^4 f^2} 
     \left[ \Big(\frac{Q^2}{s^2} + \frac{2M}{s} + s^2\Big) \partial_v \beta_i
     + \frac{2Q}{s^2}\,\partial_i Q \right]
     + \frac{2 Q b}{s^5 f^2}\right\},
\eeq
where we have integrated from $\infty$ to $r$ to derive the solution $j_i$ from (\ref{jeq}) \cite{Sin}. At the asymptotic infinity where $f \to 1$, the closed form for $j_i$ expression can be written down (where the $i$ index on the right hand side is supressed):
\beq \label{jasympt}
j_i \stackrel{r \to \infty}{\rightarrow} 
 \frac{\partial_v \beta_i}{r} + \frac{b_4}{r^4} + O\Big(\frac{1}{r^5}\Big),
\eeq
where $b_4 = \frac{M}{2}\, \partial_v \beta_i+ \frac{b Q}{2}$. This is consistent with the normalizable boundary condition which demands vanishing $r^0$ coefficient, and the requirement of the Landau frame choice that $r^{-3}$ coefficient (i.e. the boundary expectation value corresponding to $j_i$) has to vanish. Finally, $b$ is fixed by the regularity condition on the horizon,
\beq
b = -\frac{1}{r_H} \Big(1-\frac{2Q^2}{3M}\frac{1}{r_H}\Big)(4Q \partial_v \beta_i + 2 \partial_i Q),
\eeq
and $j_i (r=r_H)$ is
\beq
j_i (r_H)= \frac{1}{r_H} \Big(1-\frac{4Q^2}{3M}\frac{1}{r_H} \Big)\partial_v \beta_i
- \frac{2Q}{3M}\frac{1}{r_H^2}\,\partial_i Q
\eeq
Next, $a_i (r \to \infty)$ is obtained from integrating equation (\ref{vmodeeom1}) from infinity to $r$:
\bea
a_i (r \to \infty) &=& \frac{1}{2Q} \int_{\infty}^r ds
 \left\{ (s^4 j_i')' + 2 s \partial_v \beta_i\right\} 
= \frac{1}{2Q} [s^4 j_i'+ s^2 \partial_v \beta_i]_{\infty}^r \nn \\
&=& \frac{1}{2Q} \left\{r^4 \Big(-\frac{\partial_v \beta_i}{r^2} - \frac{4 b_4}{r^5}\Big)+ r^2 \partial_v \beta_i \right\} \nn \\
&=& -\frac{1}{r}\, \frac{2}{Q}\, b_4. \label{aiasympt}
\eea

With the standard AdS/CFT correspondence, one can derive the boundary current from the bulk solution by
\bea
\left\langle J^i \right\rangle &=& \left.\lim_{r\to\infty} \frac{\delta S}{\delta A_i}\right|_{\lambda^0}
= \lim_{r\to\infty} \sqrt{-g}\, F^{ri}
= E_i + \frac{2}{Q}\, b_4 \nn \\
&=& \left( 1- \frac{4Q^2}{3M} \frac{1}{r_H}\right)^2 E_i 
+ \frac{2}{r_H} \left(1-\frac{2Q^2}{3M}\frac{1}{r_H} \right)
\left(\frac{2Q}{3M}\, \partial_i M - \partial_i Q \right).
\label{current}
\eea
This equation exactly corresponds to the current of the fluid
\beq \label{thermocurrent}
J^i = \sigma \left( E_i - T\partial_i \frac{\mu}{T}\right),
\eeq
if we identify the temperature $T$, chemical potential $\mu$ as in (\ref{T}), and the electric conductivity $\sigma$, thermal conductivity $\kappa$ as
\bea
\sigma &=& \left( 1- \frac{4Q^2}{3M} \frac{1}{r_H}\right)^2 = \left( \frac{4\pi r_H^2 T}{3M}\right)^2 ,\nn\\
\kappa &=& \sigma T \, 
= \, \frac{1}{4\pi} \left( 1- \frac{4Q^2}{3M} \frac{1}{r_H}\right)^2 
\left( \frac{3M}{r_H^2}-\frac{4Q^2}{r_H^3}\right). \label{condsigma}
\eea
The thermodynamic relation of the current for the fluid in (\ref{thermocurrent}) is reproduced from gravity. Note that in $\sigma$, $T$ is also a function of $r_H$ and $M$, and it is straight forward to conclude from the above expression that $\sigma \to 1$ as $T \to \infty$ and $\sigma \to 0$ as $T \to 0$. The temperature and $r_H$ dependence in $\sigma$ is exactly the same as that derived in \cite{Jain} by Kubo's formula.

\begin{figure}[tb]
\begin{center}
\includegraphics[width=0.45\textwidth]{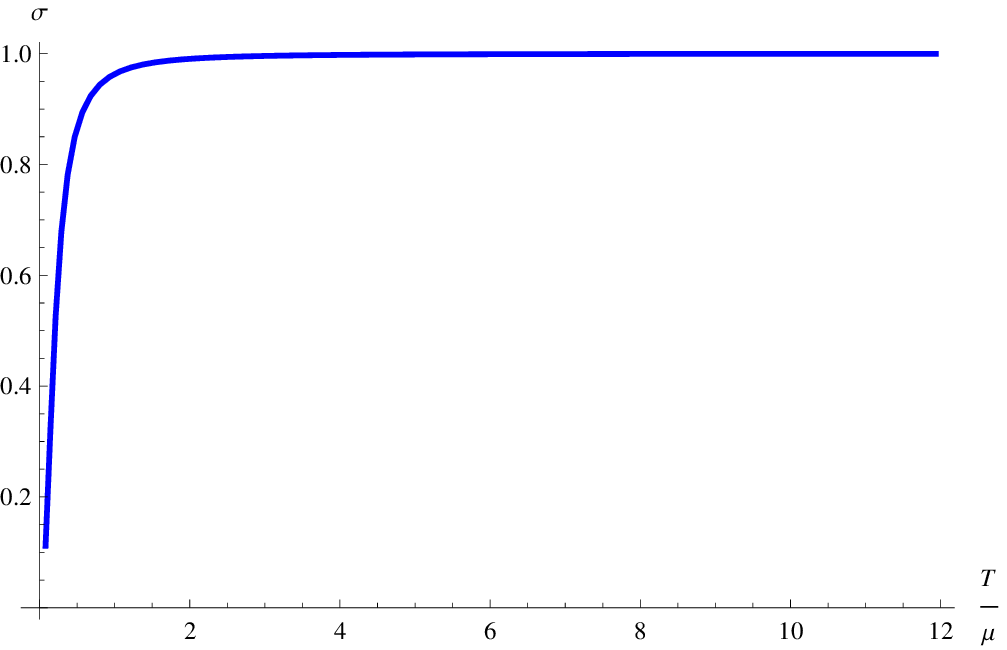}\quad\;
\includegraphics[width=0.45\textwidth]{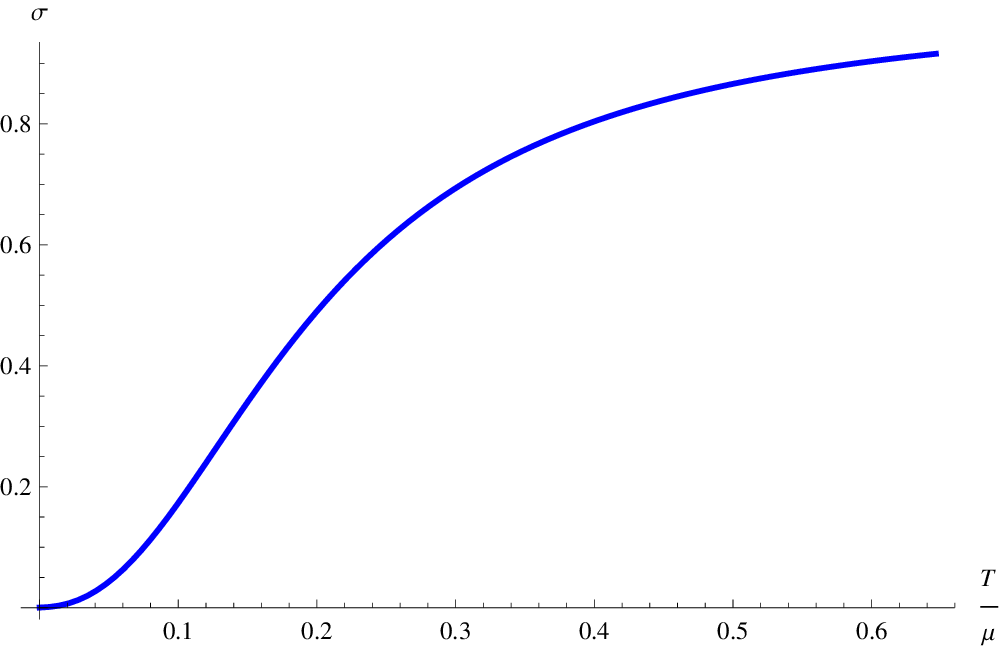}\\
\includegraphics[width=0.45\textwidth]{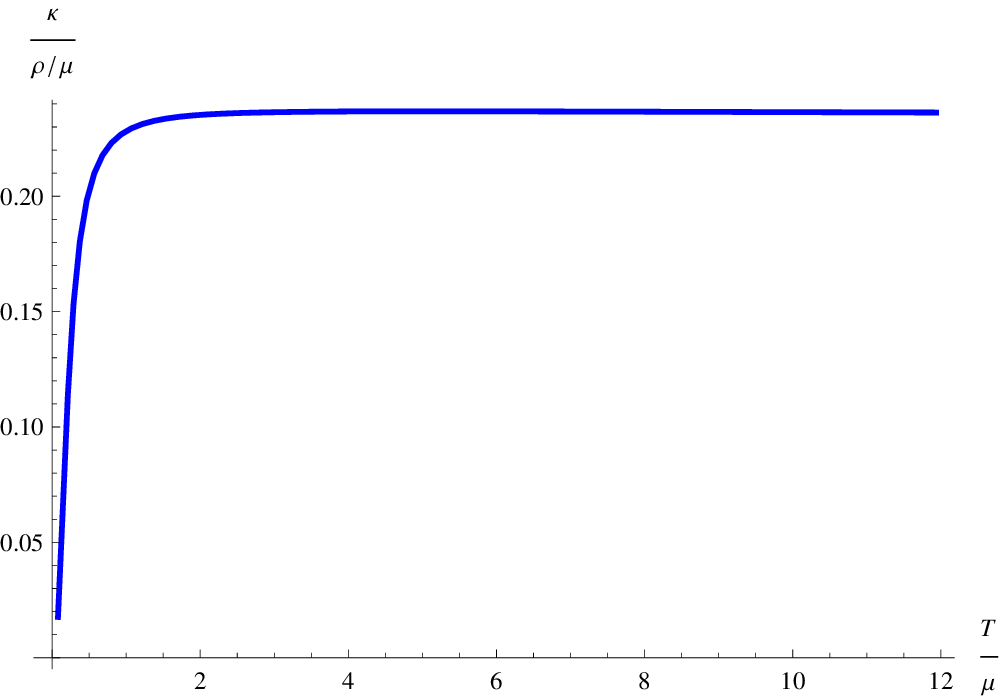}\quad\;
\includegraphics[width=0.45\textwidth]{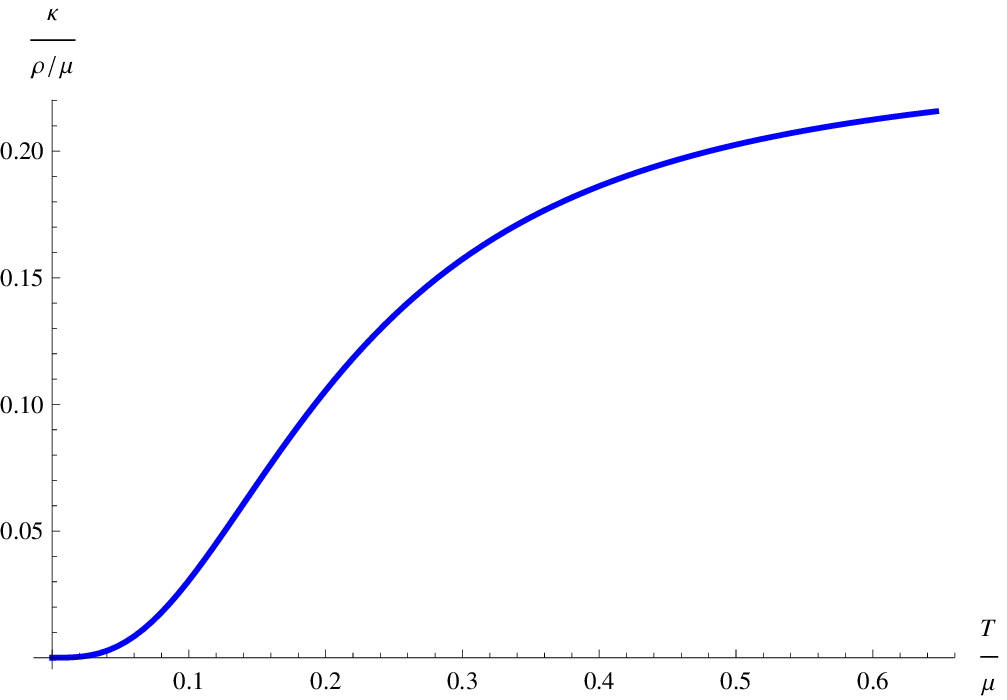}
\caption{$\sigma$ (the top figures) and $\frac{\kappa}{\rho/\mu}$ (the bottom ones) versus $\frac{T}{\mu}$, for the ranges of high $\frac{T}{\mu}$ (on the left) and low $\frac{T}{\mu}$ (on the right). $\kappa$ and $\sigma$ are related by $\kappa=\sigma T$. The value of $\sigma$ and $\frac{\kappa}{\rho/\mu}$ depend only on the charged black brane background but not on $\theta$.} \label{fig:sigma}
\end{center}
\end{figure}

The numerical behaviors of dimensionless $\sigma$ and $\frac{\kappa}{\rho/\mu}$ as functions of $T/\mu$ are displayed in Fig. \ref{fig:sigma}. This figure shows that under fixed chemical potential, the electric conductivity decreases as the temperature is reduced which is different from the usual metal property. Eventually, $\sigma$ drops to zero at $T=0$. This agrees with the result in Fig. 1 of \cite{Sin} in 3+1 dimensions. It turns out that our model predicts an ``insulator'' at zero temperature. However, this is not a conventional insulator. In general, the insulator is gapped, i.e. the conductivity is effectively zero when the temperature scale is below the energy gap. However, the smooth behavior of the conductivity reducing to zero as $T/\mu \to 0$ in the top right plot of Fig. \ref{fig:sigma} indicates that our model describes an exotic ungapped insulator at zero temperature. Since in our model, $\sigma$ depend only on the charged black brane background but not on $\theta$, the zero temperature behavior has nothing to do with the break down of the pseudo scalar probe limit. \cite{Iqbal:2010eh} demonstrates that the scalar sector undergoes a smooth phase transition at $T=0$, and according to our Fig. \ref{fig:sigma}, it turns out such phase transition related to the black brane background is also smooth and continuous. Fig. \ref{fig:sigma} also displays different low temperature behavior of $\sigma$ compared to that in \cite{Iqbal:2008by}, in which $\sigma$ is temperature independent in 2+1 dimensions. This is because our $g_{vv}$ has a second order zero at zero temperature, while the analysis in \cite{Iqbal:2008by} only applies to $g_{vv}$ with a first order zero.

On the other hand, with $\rho =2Q$, at high temperature $\sigma \to 1$ and $\frac{\kappa}{\rho/\mu} \to \frac{3}{4\pi}$. This is different from the results obtained via linear response formalism in \cite{Kovtun:2008kx} where the electric conductivity in 2+1 dimensions is temperature independent, because the gauge fields in their analysis are in the probe limit, while our background is charged and already includes the backreaction. Our result match theirs at $Q \to 0$.


\subsubsection{Parity-odd part}

The equations (\ref{vmodeeom1}) and (\ref{vmodeeom2}) implies that $j_i (r)$ and $a_i (r)$ take the form
\bea
j_i &=& j_i^{(\tilde{0})} + \lambda^2 j_i^{(\tilde{2})} + O(\lambda^3), \label{jafull}\\
a_i &=& a_i^{(\tilde{0})} + \lambda^2 a_i^{(\tilde{2})} + O(\lambda^3), \nn
\eea

The full vector mode equations (\ref{vmode1}) and (\ref{vmode2}) in the Appendix show that, in the probe limit, the leading order $O(\lambda^0)$ equations satisfied by $j_i^{(\tilde{0})}$, $a_i^{(\tilde{0})}$ as in (\ref{vmodeeom1}), (\ref{vmodeeom2}) describe only parity-even dynamics,.  The subleading $O(\lambda^2)$ equations satisfied by $j_i^{(\tilde{2})}$ and $a_i^{(\tilde{2})}$ are composed of the parity-even and parity-odd parts. The former is regarded as the higher order correction to the leading order parity-even physics. The latter, however, gives rise to the leading order parity-odd effect in the vector mode. 

As a result, $j_i^{(\tilde{2})}$, $a_i^{(\tilde{2})}$ can further be decomposed into $j_i^{(\tilde{2})} = j_{i(even)}^{(\tilde{2})} + j_{i(odd)}^{(\tilde{2})}$ and $a_i^{(\tilde{2})} = a_{i(even)}^{(\tilde{2})} + a_{i(odd)}^{(\tilde{2})}$, where the subscripts ${}_{(even)}$ and ${}_{(odd)}$ label the parity. In order to study the parity violating hydrodynamics on the boundary, we have to solve the $O(\lambda^2)_{(odd)}$ bulk equations. We will denote $j_{i(odd)}^{(\tilde{2})}$, $a_{i(odd)}^{(\tilde{2})}$ by $\bar{j_i}^{(\tilde{2})}$, $\bar{a_i}^{(\tilde{2})}$ in the following.

The two $O(\lambda^2)_{(odd)}$ equations of motion arising from $E_{ri}^{(1,\tilde{2})}$ and $r^2 M^i{}^{(1,\tilde{2})}$, after the constraint (\ref{vconsprobe}) is applied, are
\bea
\frac{1}{2r^2} \frac{d}{dr}\big( r^4 \bar{j_i}^{(\tilde{2})}{}' \big)
- \frac{1}{2} A' \bar{a_i}^{(\tilde{2})}{}'
&=& 
- \epsilon_{ij}\,\partial_j Q\,S_Q 
+ \epsilon_{ij}\,\partial_j M 
  \left[ \frac{1}{3M}\frac{1}{4r^2} \frac{d}{dr}(r^4 f' \theta') - S_M  \right] \nn \\
&& 
- \epsilon_{ij}\,E_j \Big(\frac{2Q}{3M}\Big)\frac{1}{4r^2} \frac{d}{dr}(r^4 f' \theta') 
:= \frac{S_1}{2 r^2} \, , 
\label{oddv1}
\\
\frac{d}{dr}\big( r^2 f \bar{a_i}^{(\tilde{2})}{}' \big)
- r^2 A' \bar{j_i}^{(\tilde{2})}{}'
&=&
- 4 \, g_Q' \epsilon_{ij}\,\partial_j Q
- \frac{4Q}{3M}\, g_M' \epsilon_{ij}\,\partial_j M 
+ 2 \, g_E' \epsilon_{ij}\,E_j \nn \\
&:=& S_2 \, ,
\label{oddv2}
\eea
where $\frac{S_1}{2r^2}$ and $S_2$ denote the sources of these differential equations, and
\bea
S_M &=& \frac{1}{2 r^6 f'} \left(r^7 f'^2\right)' \frac{\partial \theta}{\partial M}
 -\frac{1}{4} \left(r^2 f'\right)' \frac{\partial \theta'}{\partial M} 
 +\frac{1}{4r^2} \left(r^4 \theta' \frac{\partial f'}{\partial M} \right)' \, , \nn 
\\
S_Q &=& \frac{1}{2 r^6 f'} \left(r^7 f'^2\right)' \frac{\partial \theta}{\partial Q}
 -\frac{1}{4} \left(r^2 f'\right)' \frac{\partial \theta'}{\partial Q} 
 +\frac{1}{4r^2} \left(r^4 \theta' \frac{\partial f'}{\partial Q} \right)' \, ,\nn
\\ 
g_Q'(r)&=& \frac{\theta'}{r} + \frac{Q}{r^2}\frac{\partial \theta}{\partial Q}\, , \label{beta}\\
g_M'(r)&=& -\frac{\theta'}{r}+\frac{3M}{r^2}\frac{\partial \theta}{\partial M}, \nn \\
g_E'(r)&=& \Big(1-\frac{4Q^2}{3M}\frac{1}{r}\Big)\theta' \, .\nn
\eea
It is clear from above that the functions $g_Q$, $g_M$, $g_E$ are no more dominant than $O(\frac{1}{r^{\Delta_-}})$ as $r \to \infty$. 

Integrating (\ref{oddv2}) from $r_H$ to $r$ on both sides and then substituting $\bar{a}_i^{(\tilde{2})}$ into (\ref{oddv1}), one arrives at the differential equation for $\bar{j_i}^{(\tilde{2})}$:
\beq 
\frac{d}{dr}\Big\{ r^4 f^2 \frac{d}{dr}\Big(\frac{\bar{j_i}^{(\tilde{2})}(r)}{f}\Big)
\Big\} 
=
-\frac{4Q^2}{r^2} \bar{j_i}^{(\tilde{2})}(r_H) + f\,S_1 
+ \frac{2Q}{r^2} \int_{r_H}^r S_2
\eeq
The solution is obtained by integrating the above equation from $\infty$ to $r$ on both sides twice,
\beq
\bar{j_i}^{(\tilde{2})} (r)
= f \int_{\infty}^r \frac{1}{s^4 f^2}
  \left\{\frac{4Q^2}{s}\bar{j_i}^{(\tilde{2})}(r_H)
 +\int_{\infty}^s \left[ f\,S_1 + \frac{2Q}{u^2}\int_{r_H}^u S_2 \right]
\right\}\,.
\eeq
Then we can write down the expression of $\bar{j_i}^{(\tilde{2})}$ on the horizon:
\beq \label{jbarrH}
\bar{j_i}^{(\tilde{2})} (r_H) 
= -\frac{1}{3M}\int_{\infty}^{r_H}
  \left\{f\, S_1 + \frac{2Q}{r}\int_{r_H}^r S_2 \right\} \, .
\eeq
This expression is consistent with the regularity condition. On the other hand, careful analysis shows that $\bar{j_i}^{(\tilde{2})}$ behaves asymptotically as 
\beq \label{jbarasympt}
\bar{j_i}^{(\tilde{2})} (r\to\infty) \sim
\frac{\bar{b}_4}{r^4}
+ O(\frac{1}{r^{\Delta_- + 4}}) + \cdots\, ,
\eeq
where $\cdots$ are terms of higher order than $O(\frac{1}{r^{\Delta_- + 4}})$, and
\beq \label{j4bar}
\bar{b}_4 = -Q^2 \bar{j_i}^{(\tilde{2})}(r_H)\,.
\eeq
Since we limit ourself to $ 0 < \Delta_- < \frac{3}{2}$, 
the leading order in (\ref{jbarasympt}) is $r^{-4}$. Moreover, (\ref{jbarasympt}) is consistent with the boundary conditions stated below (\ref{jasympt}). 

Then, we can evaluate the large $r$ behavior of $\bar{a}_i^{(\tilde{2})}$ by integrating (\ref{oddv1}) on both side from $\infty$ to some large $r$ as in (\ref{aiasympt}),
\beq 
\bar{a}_i^{(\tilde{2})} (r \to \infty) \sim
- \frac{1}{r} \frac{2}{Q} \bar{b}_4 + \dots.
\eeq
The coefficients of $1/r$ in $\bar{a}_i^{(\tilde{2})} (r \to \infty)$ is interpreted as the expectation value corresponding to $\bar{a}_i^{(\tilde{2})}$ on the boundary. Indeed, this 
is given by
\bea 
\left\langle \bar{J^i} \right\rangle &=& \lim_{r \to \infty} \left. \frac{\delta S^{(1,\tilde{2})}}{\delta A_i}\right|_{(odd)} = \lim_{r \to \infty} \sqrt{-g} \bar{F}^{ri}{}^{(\tilde{2})}
= \frac{2}{Q}\, \bar{b}_4  \nn \\
&=&
 \bar{K}_E\, \epsilon_{ij} E_j 
+ \bar{K}_Q \,\epsilon_{ij} \partial_j Q 
+ \bar{K}_M \,\epsilon_{ij} \partial_j M \, .\label{Jibar}
\eea 
After some algebra, one obtains
\bea
\bar{K}_E &=& -2 \left\{
\frac{r_H^4 f'|_{r_H}}{3M} g_E|_{r_H}
-\frac{4Q^2}{3M} \int_{\infty}^{r_H} \frac{g_E}{r^2} 
-\frac{Q^2}{9M^2} \int_{\infty}^{r_H} \left(r^2f'\right)^2 \theta' \right\}\, , \nn
\\
\bar{K}_Q &=& 4 \left\{
\frac{r_H^4 f'|_{r_H}}{3M} g_Q|_{r_H} 
- \frac{4Q^2}{3M} \int_{\infty}^{r_H} \frac{g_Q}{r^2} 
- \frac{Q}{3M} \int_{\infty}^{r_H} r^2 f S_Q
\right\}\, , \label{Kbar}
\\
\bar{K}_M &=& \frac{4Q}{3M} \left\{
\frac{r_H^4 f'|_{r_H}}{3M} g_M|_{r_H} 
- \frac{4Q^2}{3M} \int_{\infty}^{r_H} \frac{g_M}{r^2} 
- \int_{\infty}^{r_H} r^2 f S_M -\frac{1}{12M} 
  \int_{\infty}^{r_H} \left(r^2 f'\right)^2 \theta'
\right\}\, , \nn
\eea
in which we have carried out integration by part for the term $\int_{\infty}^{r_H} f (r^4 f' \theta')'$.

Let's compare the expression (\ref{Jibar}) with the parity violating part of the constitutive equation of the boundary current $J^{i}$ in (\ref{Jmu}):
\beq
J^{i} = \widetilde{\sigma} \epsilon_{ij}\, E_j
+ \widetilde{\kappa}\, \epsilon_{ij} \nabla_j \frac{\mu}{T}
+ \widetilde{\xi}\, \epsilon^{ij} \nabla_j T\, .
\eeq
Substituting in $T$, $\mu$ from the bulk theory in (\ref{T}), one can identify the Hall conductivity $\widetilde{\sigma}$, thermal Hall conductivity $\widetilde{\kappa}$, and heat Hall conductivity $\widetilde{\xi}$ in terms of $\bar{K}_E$, $\bar{K}_Q$ and $\bar{K}_M$:
\bea
\widetilde{\sigma} &=& \bar{K}_E,\nn\\
\widetilde{\kappa} &=& -\frac{1}{4\pi r_H^2}\frac{3Mr_H-4Q^2}{6Mr_H-4Q^2} \big(2Q\bar{K}_M + 3M r_H \bar{K}_Q\big), \label{sigmat}\\
\widetilde{\xi} &=& \frac{4\pi r_H^3}{3M r_H - 4Q^2} \big(3M \bar{K}_M+ 2Q \bar{K}_Q \big). \nn
\eea

Since $\bar{K}_M$, $\bar{K}_Q$ and $\bar{K}_M$ contain integrals of functions of $\theta$ and $\theta'$, their values are subject to the choice of the boundary condition for $\theta$. In the following we present the numerical result of $\widetilde{\sigma}$, $\widetilde{\kappa}$, and $\widetilde{\xi}$ according to the two different types of boundary conditions we choose:

\noindent{\it (i) Asymptotically sourced $\theta$:} 

\begin{figure}[tbp]
\begin{center}
\includegraphics[width=0.49\textwidth]{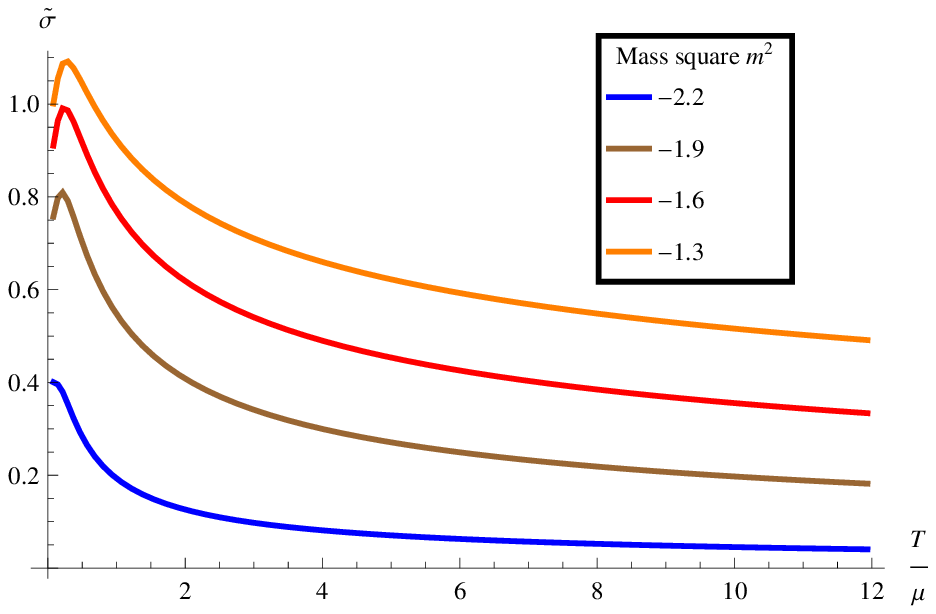}
\includegraphics[width=0.49\textwidth]{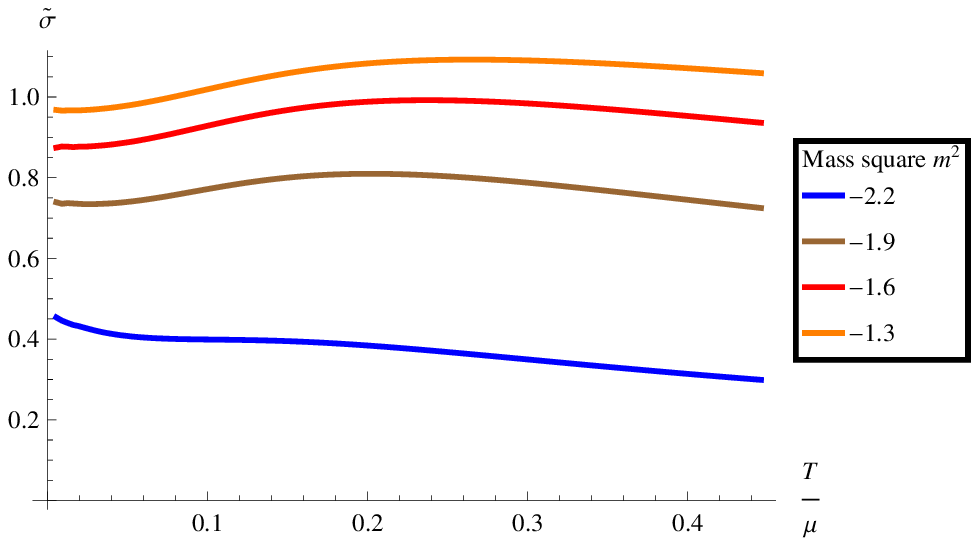}
\caption{$\widetilde{\sigma}$ versus $T/\mu$ for various $\theta$ mass $m^2$, under the sourced boundary condition for $\theta$. The right figure is for small $T/\mu$ while the left one is for a wider range of $T/\mu$.} \label{fig:AHC}
\includegraphics[width=0.49\textwidth]{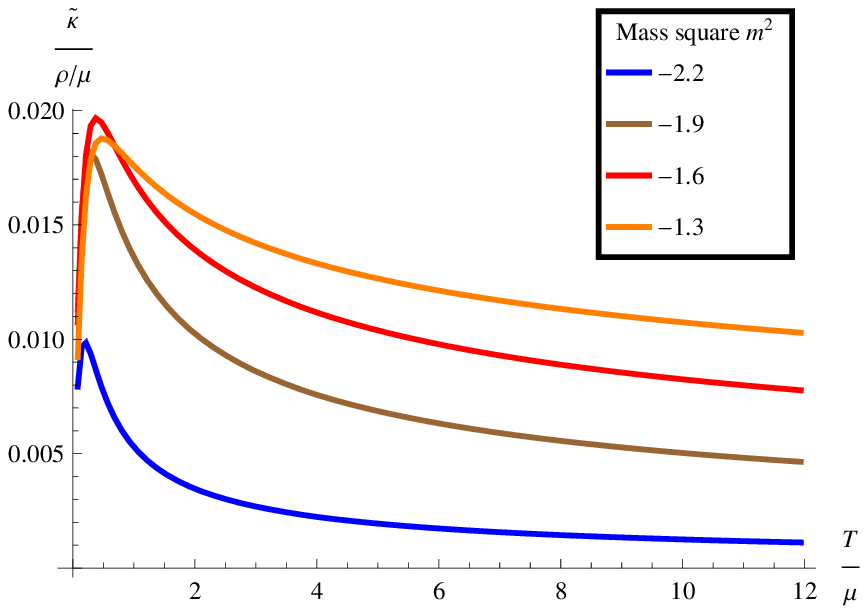}
\includegraphics[width=0.49\textwidth]{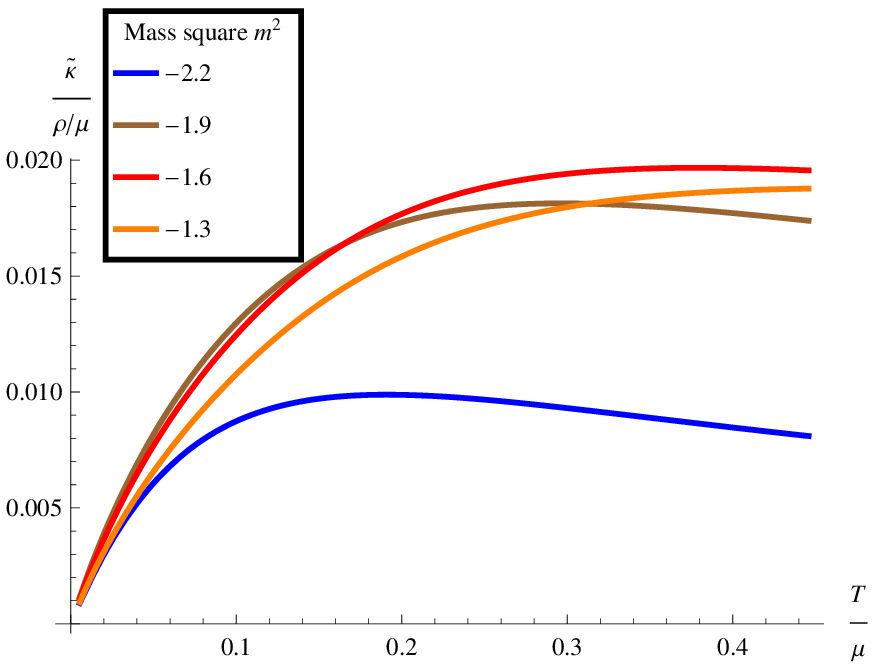}
\caption{$\frac{\widetilde{\kappa}}{\rho/\mu}$ versus $T/\mu$ for various $m^2$, under the sourced boundary condition for $\theta$. The right figure is for small $T/\mu$ while the left one is for a wider range of $T/\mu$.} \label{fig:ATHC}
\includegraphics[width=0.49\textwidth]{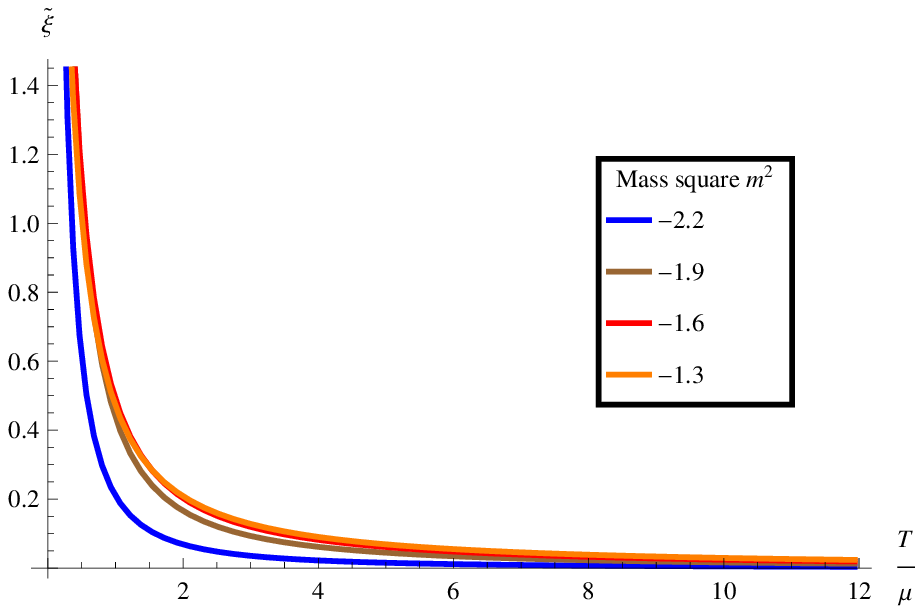}
\includegraphics[width=0.49\textwidth]{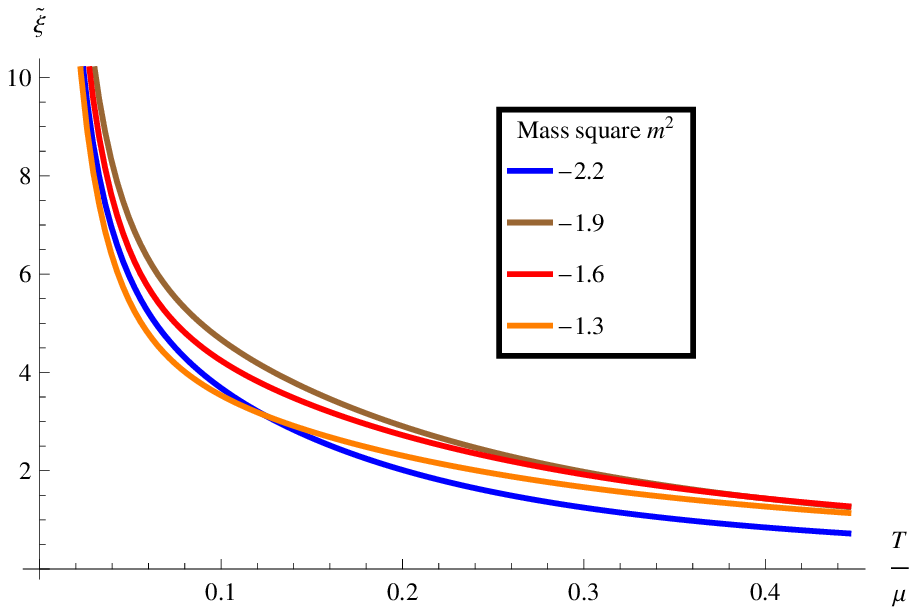}
\caption{$\widetilde{\xi}$ versus $T/\mu$ for various $m^2$, under the sourced boundary condition for $\theta$. The right figure is for small $T/\mu$while the left one is for a wider range of $T/\mu$ .} \label{fig:AHHC}
\end{center}
\end{figure}

As the source of $\theta$ on the asymptotic boundary is turned on, the numerical results summarized in Fig. \ref{fig:AHC} $\sim$ \ref{fig:AHHC} shows that: $\widetilde{\sigma}$ is positive and finite throughout who range of $T/\mu$; $\frac{\widetilde{\kappa}}{\rho/\mu}$ is also positive and increases as $T/\mu$ reduces from large value before it starts to decrease below some small $T/\mu$. It eventually vanishes at $T=0$ as predicted by (\ref{sigmat}); $\widetilde{\xi}$ is positive,  increases monotonically as $T/\mu$ reduces, and diverges at $T/\mu=0$.

\noindent{\it (ii) Asymptotically sourceless $\theta$:}

\begin{figure}[hp]
\begin{center}
\includegraphics[width=0.45\textwidth]{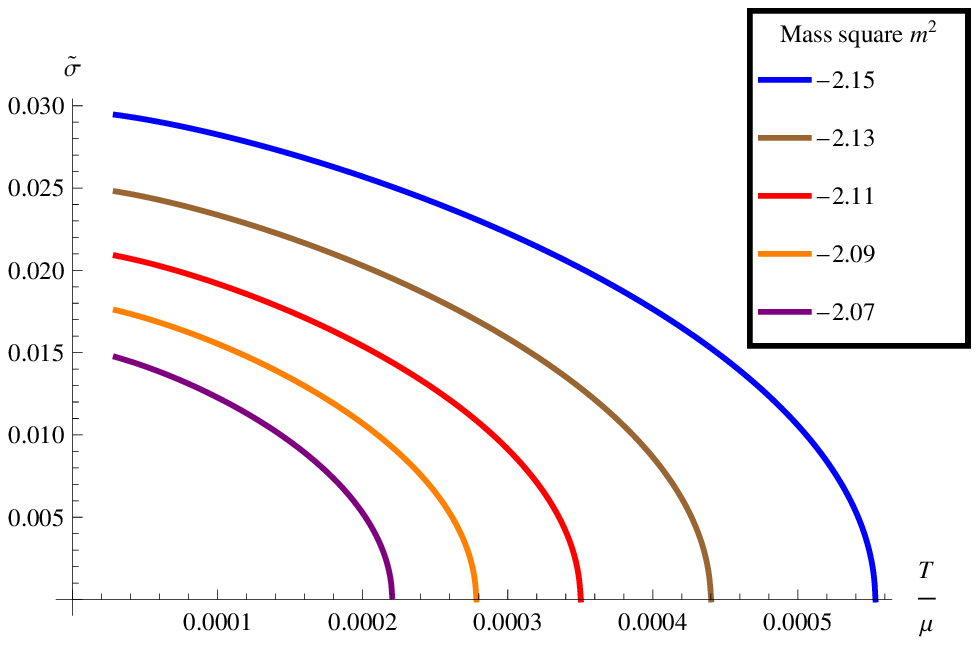}
\includegraphics[width=0.45\textwidth]{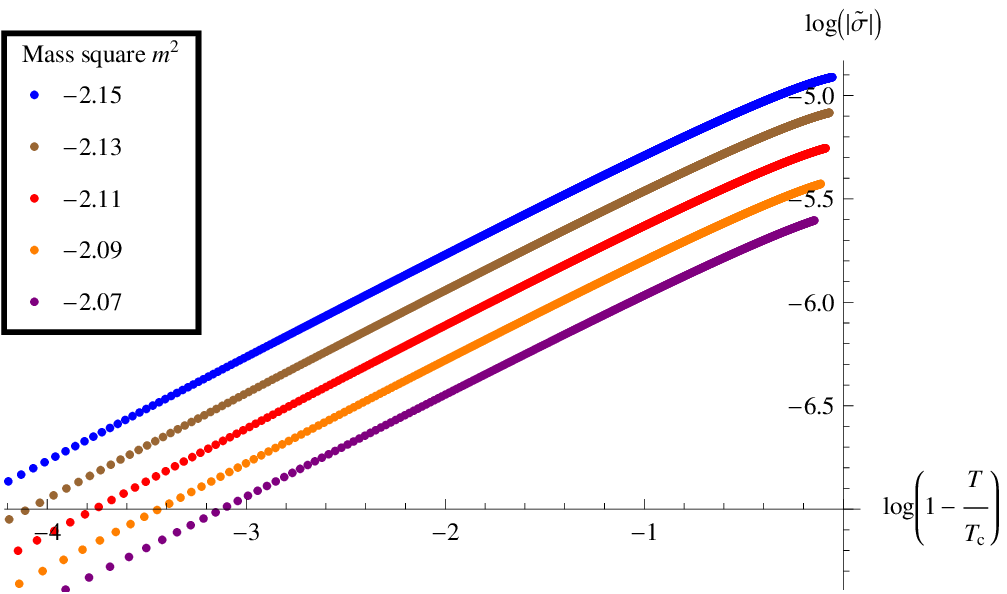}
\caption{$\widetilde{\sigma}$ versus $T/\mu$ with linear scale (on the left) and log scale (on the right) as $\theta$ is asymptotically sourceless. The left figure shows $\widetilde{\sigma}$ vanishes at and above the critical temperature which depends on $m^2$, and the right one indicates the critical exponent is $0.5$.} \label{fig:AHCns}
\includegraphics[width=0.45\textwidth]{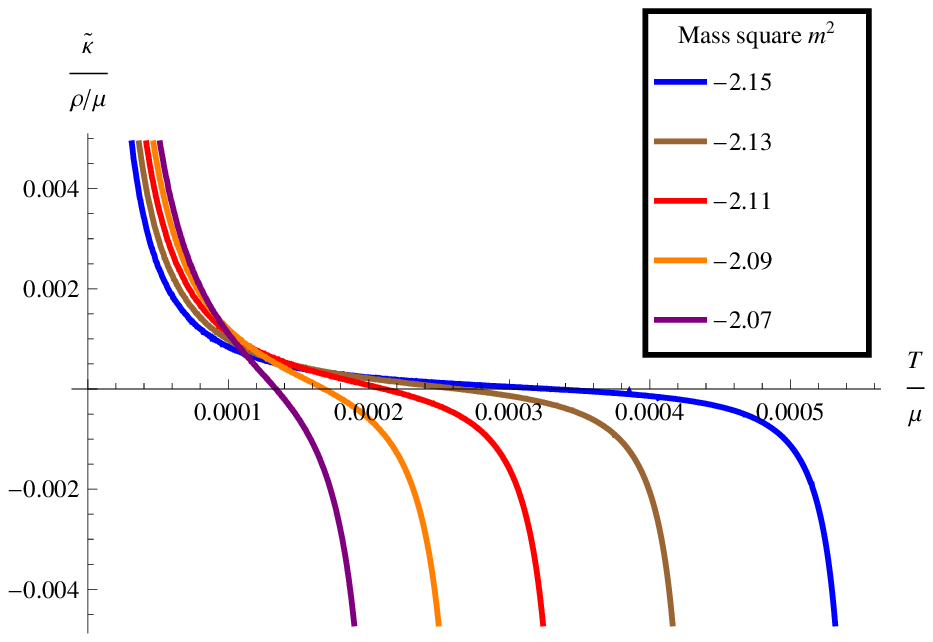}
\includegraphics[width=0.45\textwidth]{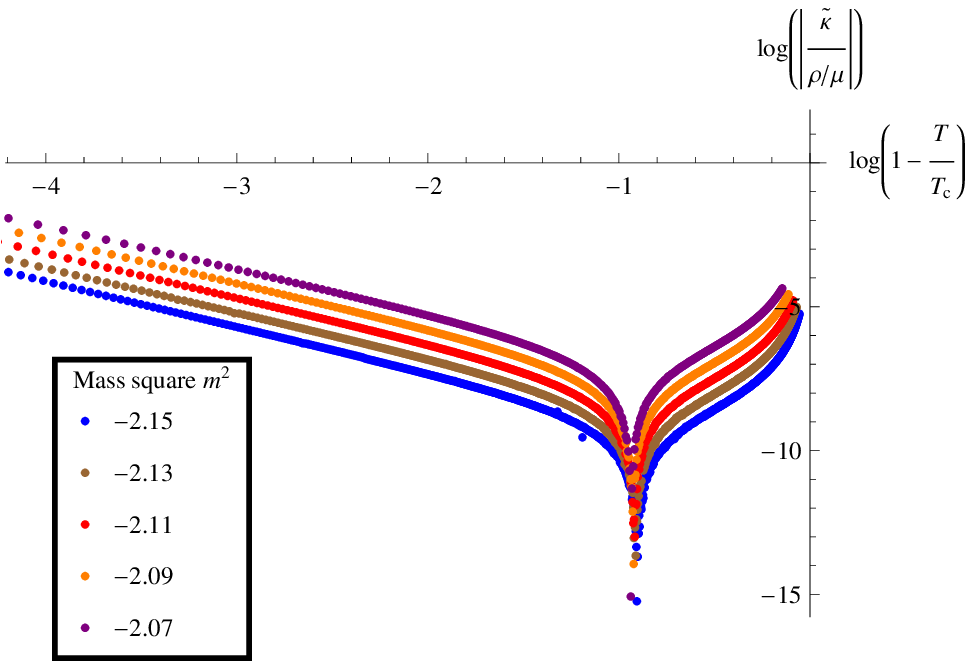}
\caption{$\frac{\widetilde{\kappa}}{\rho/\mu}$ versus $T/\mu$ with linear scale (on the left) and log scale (on the right) as $\theta$ is asymptotically sourceless. The left figure shows that $\frac{\widetilde{\kappa}}{\rho/\mu}$ diverges at the critical temperature, with a critical exponent $-1.5$ as being indicated by the right one.} \label{fig:ATHCns}
\includegraphics[width=0.49\textwidth]{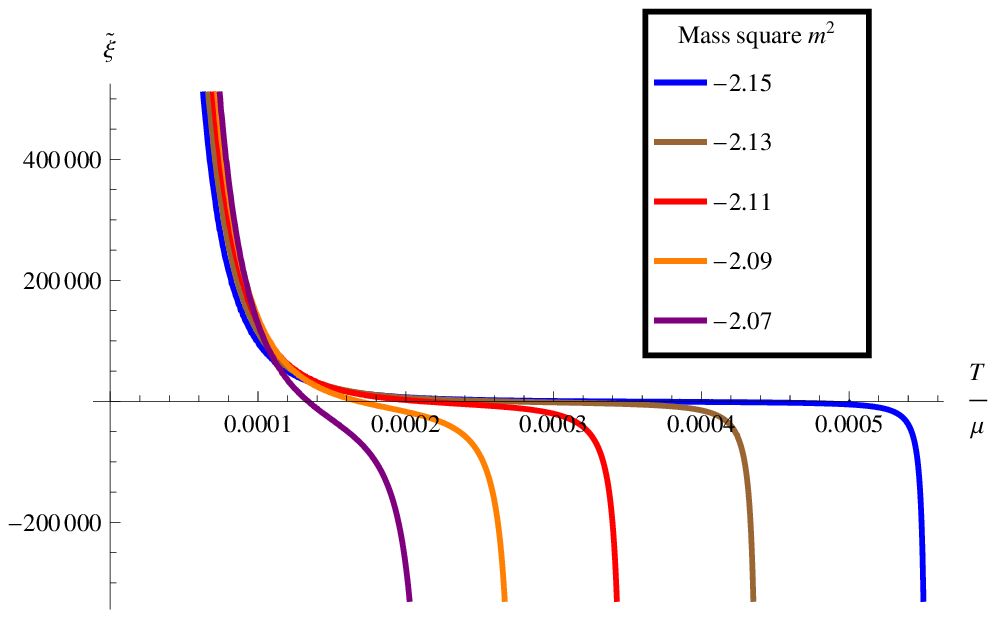}
\includegraphics[width=0.49\textwidth]{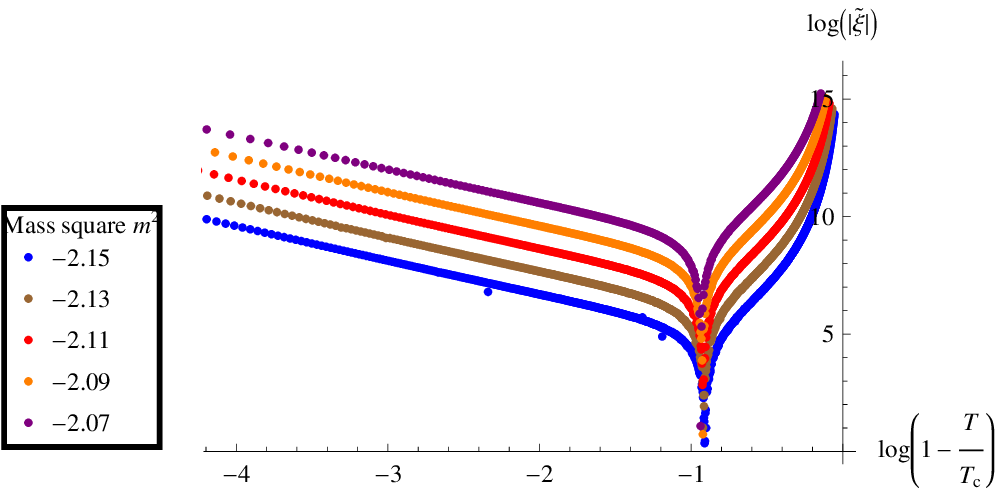}
\caption{$\widetilde{\xi}$ versus $T/\mu$ with linear scale (on the left) and log scale (on the right) as $\theta$ is asymptotically sourceless. The left figure shows that $\widetilde{\xi}$ diverges at the critical temperature, with a critical exponent $-1.5$ as being indicated by the right one.} \label{fig:AHHCns}
\end{center}
\end{figure}

For the case where $\theta$ is asymptotically sourceless and the boundary is parity spontaneously broken by the nonzero vev of the dual pseudo scalar operator, Fig. \ref{fig:AHCns}$\sim$\ref{fig:AHHCns} display the numerical results of the three parity violating  coefficients from the vector mode: $\widetilde{\sigma}$ is nonvanishing and positive below the critical temperature $T_c$, with a critical exponent $0.5$ near $T_c$. $\frac{\widetilde{\kappa}}{\rho/\mu}$ and $\widetilde{\xi}$ both diverge at $T=T_c$ and $T=0$, both with critical exponents -1.5 close to $T=T_c$. These critical exponents, however, cannot be derived from (\ref{sigmat}), as we don't have an analytical expression of $\theta$.

\subsection{Tensor mode}

For the tensor mode, there is one dynamical equation due to $H E_{xy}^{(1)}$. 
In the backreacted case it reads
\begin{equation}
\frac{d}{dr}\left[ -\frac{1}{2}\frac{r^{4}f}{H}\frac{d}{dr}
\alpha_{xy}\right] -r(\partial_{x}\beta_{y}+\partial_{y}\beta_{x})-
\frac{\lambda }{4}\frac{d}{dr}(\frac{r^{4}f^{\prime }\theta^{\prime }}{H^{2}
})(\partial_{x}\beta_{x}-\partial_{y}\beta_{y})=0.  \label{Exy1}
\end{equation}


\subsubsection{Parity-even part}

After taking the probe limit $\theta \rightarrow \lambda \theta $ and 
$\varphi \rightarrow \lambda \varphi$, the $O(\lambda^0)$ part of the above equation becomes
\begin{equation}
\frac{d}{dr}\left[ -\frac{1}{2}r^{4}f\frac{d}{dr}%
\alpha_{xy}^{(\widetilde{0})}\right] = r(\partial_{x}\beta_{y}+\partial_{y}\beta_{x}),  \label{Exy10s}
\end{equation}
in which the source only contains the shear mode. This equation can be analytically solved by integrating on both sides twice, with first integration being indefinite and the second one from $\infty$ to $r$:
\beq \label{alpha}
\alpha_{xy}^{(\widetilde{0})} = -\int_{\infty}^{r}\frac{\partial_{x}\beta_{y}+\partial_{y}\beta_{x}}{r^2 f} 
- \int_{\infty}^{r} \frac{2 \alpha_3}{r^4 f}.
\eeq
The asymptotic behavior of $\alpha_{xy}^{(\widetilde{0})}$ is
\beq
\alpha_{xy}^{(\widetilde{0})}(r\to\infty) \sim
\frac{\partial_{x}\beta_{y}+\partial_{y}\beta_{x}}{r} + \frac{2 \alpha_3}{3r^3} + O(\frac{1}{r^4}),
\eeq
which is consistent with the sourceless boundary condition that the $O(r^0)$ coefficient vanishes. The $O(r^{-3})$ coefficient gives rise to the expectation value, 
\beq \label{vevT}
\left\langle T_{xy}^{(\widetilde{0})} \right\rangle = \frac{3}{16\pi G_N} \frac{2}{3}\alpha_3, 
\eeq
from which the shear viscosity $\eta$ can be extracted. The constant $\alpha_3$ is determined by the regularity condition on the horizon for $\alpha_{xy}^{(\widetilde{0})}$
\beq \label{regalpha}
\frac{d}{dr}\alpha_{xy}^{(\widetilde{0})}(r_{H})
= -\frac{2(\partial_{x}\beta_{y}+\partial_{y}\beta_{x})}{r_H^3 f'(r_H)} 
\eeq
and (\ref{alpha}):
\beq
\alpha_3 = \left(\frac{r_H f(r_H)}{f'(r_H)}-\frac{r_H^2}{2}\right) (\partial_{x}\beta_{y}+\partial_{y}\beta_{x}).
\eeq
Then one immediately obtains the shear viscosity,
\beq
\eta = \frac{r_H^2}{16\pi G_N}.
\eeq
With the entropy density $s=\frac{r_H^2}{4G_N}$, the $\eta/s$ ratio match the universal bound,  
\beq
\frac{\eta }{s}= \frac{1}{4\pi}.
\eeq


\subsubsection{Parity-odd part}

Next, we proceed to $O(\lambda^2)$ order of the tensor mode equation (\ref{Exy1}). For this purpose we consider the $\alpha_{xy}$ ansatz up to $O(\lambda^2)$,
\beq
\alpha_{xy} = \alpha_{xy}^{(\widetilde{0})}+\lambda^2 \alpha_{xy}^{(\widetilde{2})},
\eeq
and substitute to (\ref{Exy1}). Then it is easy to see that the $O(\lambda^2)$ equation is composed of parity even and parity odd part, which implies that $\alpha_{xy}^{(\widetilde{2})}$ can further be decomposed into two independent modes
\beq
\alpha _{xy}^{(\widetilde{2})}=\alpha _{xy}^{(\widetilde{2})}(r)(\partial
_{x}\beta _{y}+\partial _{y}\beta _{x})+\bar{\alpha} _{xy}^{(\widetilde{2}%
)}(r)(\partial _{x}\beta _{x}-\partial _{y}\beta _{y}),
\eeq
where the shear mode is parity even and regarded as the $O(\lambda^2)$ correction of the shear viscosity, while the Hall mode (proportional to  $\partial_x \beta_x - \partial_y \beta_y$) is the leading order parity odd effect. In the following we focus on the parity violating mode, $\alpha _{xy}^{(\widetilde{2})}{}_{(odd)} = \bar{\alpha} _{xy}^{(\widetilde{2})}(r)(\partial _{x}\beta _{x}-\partial _{y}\beta _{y})$ and extract the Hall viscosity from bulk gravity.

The $O(\lambda^2)$ parity odd equation is
\beq
\frac{d}{dr}\left[ -\frac{1}{2}r^{4}f\frac{d}{dr}
\bar{\alpha}_{xy}^{(\widetilde{2})}\right]
= -\frac{1}{4}\frac{d}{dr} (r^4 f' \theta')
\eeq
Integrating both sides twice, with first integration being indefinite and the second one from $\infty$ to $r$, one can write down the analytic solution for $\bar{\alpha}_{xy}^{(\widetilde{2})}$,
\beq \label{alphabar}
\bar{\alpha}_{xy}^{(\widetilde{2})} (r)
=  \int_{\infty}^r \frac{f' \theta'}{2f}
 - \int_{\infty}^r \frac{2 \bar{\alpha_3}}{r^4 f}.
\eeq
The asymptotic behavior is 
\beq
\bar{\alpha}_{xy}^{(\widetilde{2})} (r \to \infty) \sim
\frac{2 \bar{\alpha}_3 }{3r^3} + O(\frac{1}{r^{\Delta_- + 4}}).
\eeq
This is consistent with the normalizable boundary condition. According to (\ref{vevT}), the Hall viscosity $\widetilde{\eta}_A$ is encoded in $\bar{\alpha}_3$ which is determined by demanding (\ref{alphabar}) satisfy the regularity condition on the horizon,
\beq
\frac{d}{dr}\bar{\alpha}_{xy}^{(\widetilde{2})}(r_{H})
= \left. -\frac{1}{2} \frac{1}{r^{4}f'} \frac{d}{dr}(r^{4}f' \theta')\right|_{r=r_{H}}.
\eeq
Then $\bar{\alpha}_3$ is fixed to
\beq
\bar{\alpha}_3 = \left.\frac{f}{4f'}\big(r^4 f' \theta'\big)'\right|_{r_H}
+ \left( \frac{3M}{4}-\frac{Q^2}{r_H}\right)\theta'(r_H),
\eeq
\begin{figure}[tbp]
\begin{center}
\includegraphics[width=0.49\textwidth]{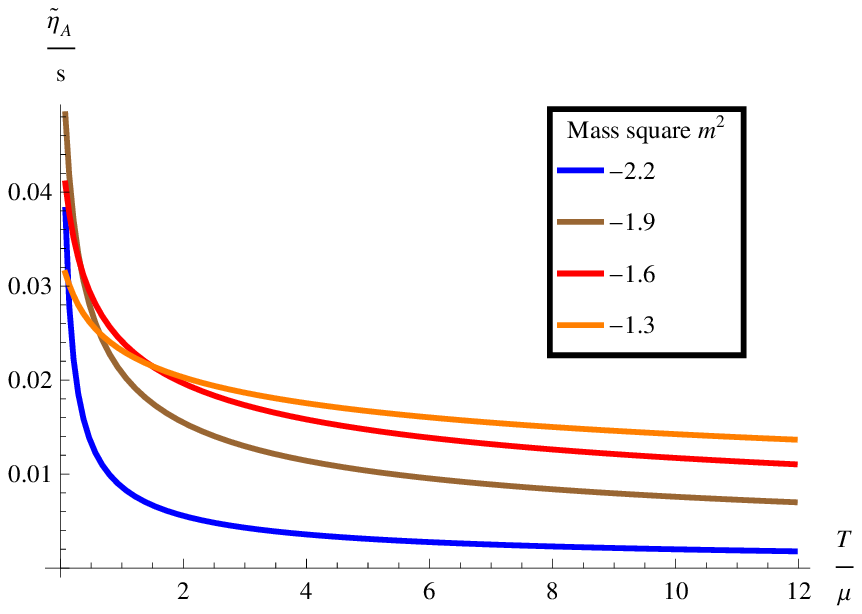}
\includegraphics[width=0.49\textwidth]{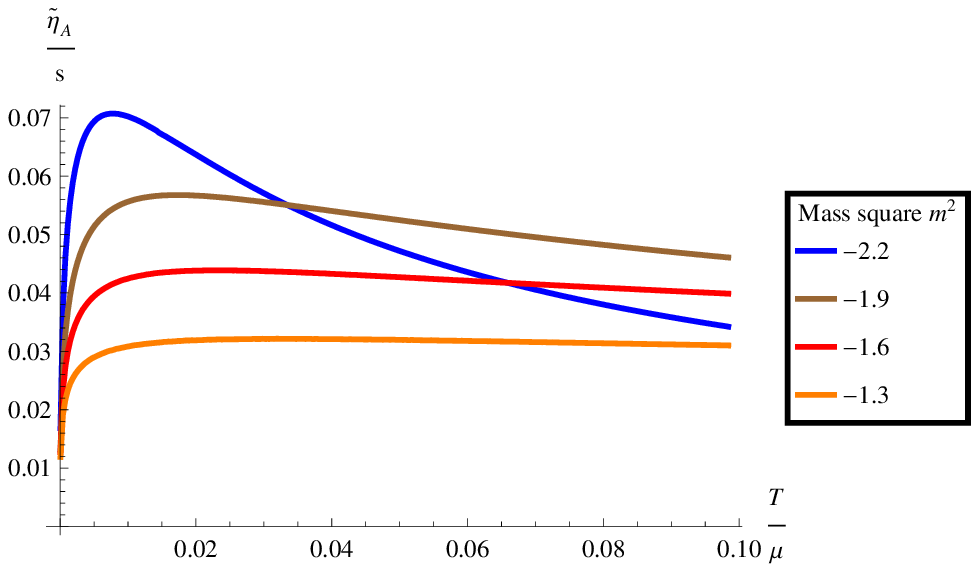}
\caption{$\widetilde{\eta}_A/s$ versus $T/\mu$ for asymptotically sourced $\theta$. The right figure is for small $T/\mu$, while the left one for a wider range of $T/\mu$.} \label{fig:HallV}
\end{center}
\end{figure}
such that \cite{son2}
\bea
\widetilde{\eta}_A &=& -\frac{r_H^4 f'(r_H) \theta'(r_H)}{32\pi G_N} 
       = - \frac{r_H^2}{32\pi G_N} \left.\frac{dV}{d\theta}\right|_{r_H} 
       = - \left.\frac{r_H^2}{32\pi G_N} \big( m^2 \theta + c \theta^3\big)\right|_{r_H}, \label{etaA} \\ 
\frac{\widetilde{\eta}_A}{s} &=& -\left.\frac{1}{8\pi}\frac{dV}{d\theta}\right|_{r_H}
=-\left.\frac{1}{8\pi}\big( m^2 \theta + c \theta^3\big)\right|_{r_H}. \nn
\eea
In the above expressions we have applied the regularity condition on the horizon for $\theta$ in (\ref{thetareg}). It is clear from (\ref{etaA}) that the Hall viscosity is controlled by the value of $\theta$ on the horizon as the mass square of the pseudo scalar. Since $\theta(r_H)$ is affected by the choice of the asymptotic boundary condition, so is Hall viscosity:

\noindent{\it (i) Asymptotically sourced $\theta$:}

Fig. \ref{fig:HallV} displays $\widetilde{\eta}_A/s$ as a function of $T/\mu$, as $\theta$ is subject to the condition that the asymptotic source on the boundary is turned on. The numerical analysis shows that, as the temperature reduces, $\widetilde{\eta}_A/s$ increases monotonically until the temperature is below some small value, and then $\widetilde{\eta}_A/s$ decreases along with the temperature and eventually reaches some finite  value at $T=0$. 

\noindent{\it (ii) Asymptotically sourceless $\theta$:}

\begin{figure}[tbp]
\begin{center}
\includegraphics[width=0.49\textwidth]{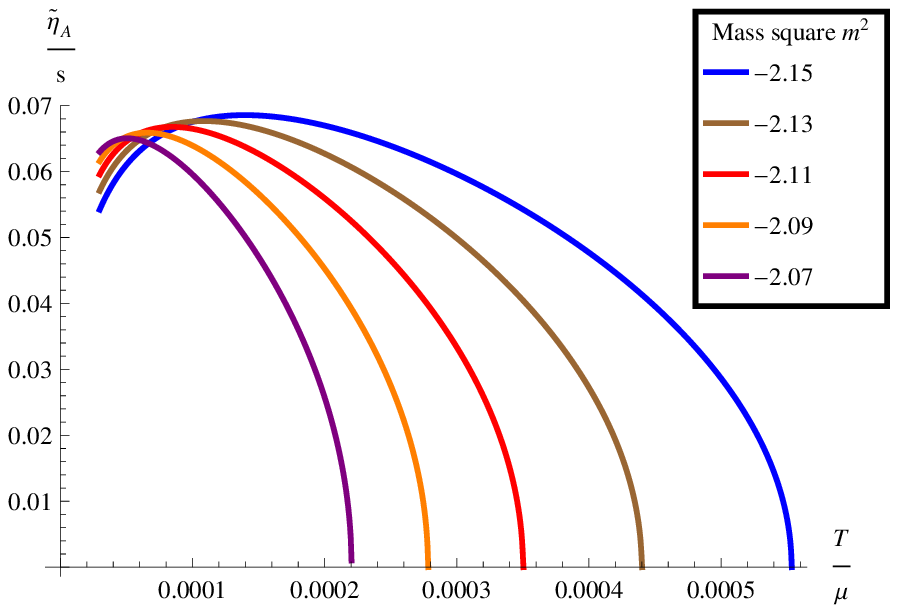}
\includegraphics[width=0.49\textwidth]{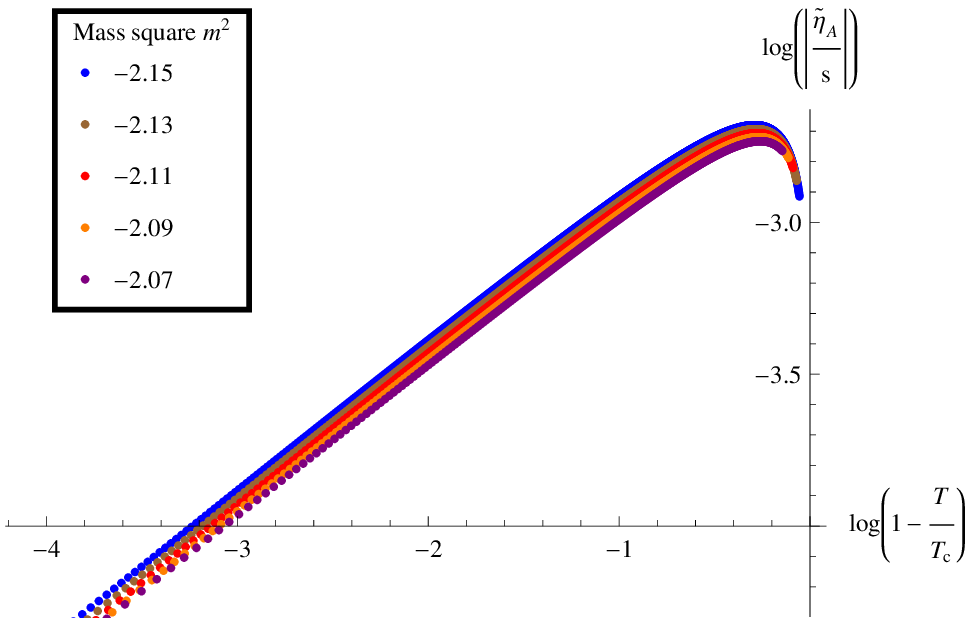}
\caption{$\widetilde{\eta}_A/s$ versus $T/\mu$ with linear scale (on the left) and log scale (on the right) for various $\theta$ mass $m^2$, as $\theta$ is asymptotically sourceless. The left figure shows the Hall viscosity vanishes at and above some critical temperature, and the right figure indicates the critical exponent equal to $0.5$.} \label{fig:HallVns}
\end{center}
\end{figure}

Fig. \ref{fig:HallVns} shows the numerical results of $\widetilde{\eta}_A/s$ versus $T/\mu$ when $\theta$ is sourceless on the boundary. In this case, the Hall viscosity acquires non-vanishing value only below the critical temperature $T_c$ depending on $m^2$. The critical exponent near $T_c$ is $0.5$, as expected from the analysis of $\theta$.

We close this subsection by recalling the analytic solution of $\alpha_{xy}$ for the general case with backreaction studied in \cite{Chen}:
\beq \label{fullalpha}
\alpha_{xy}(r)=\int_{r}^{\infty} \frac{2H(t)dt}{t^{4}f(t)}\int_{r_{H}}^{t}dz
\left[ z(\partial_{x}\beta_{y}+\partial_{y}\beta_{x})
+\frac{\lambda }{4} \frac{d}{dz}(\frac{r^{4}f'\theta'}{H^{2}})
(\partial_{x}\beta_{x} - \partial_{y}\beta_{y})\right] .
\eeq
The above results of the shear and Hall viscosities are consistent with those derived from (\ref{fullalpha}) in \cite{Chen}.

\section{Discussion and conclusion} \label{conclusion}

We have constructed a 3+1 dimensional holographic model with a charged black brane and neutral pseudo scalar field $\theta$, dual to the parity violating hydrodynamic system in 2+1 dimensions. We have extracted all the transport coefficients from the scalar, vector, and tensor modes of bulk perturbations, and reconstructed the hydrodynamics to first order in derivatives in the probe limit of $\theta$. In our model, the parity violating effect is contributed from the neutral pseudo scalar $\theta$, coupled to the gravitational and electrodynamical Chern-Simons terms. 

The potential term in (\ref{V}) allows the parity of the system to be broken spontaneously by the non-trivial vev of the boundary operator dual to $\theta$ developed below $T_c$, such that all parity transport coefficients are the order parameter for the second order phase transition and are switched on below $T_c$. They exhibit certain critical exponent behavior very close to $T_c$. Alternatively, one can choose the sourced boundary condition at asymptotic infinity, such that the parity and the conformal symmetry of the system is broken by this external source. The finite temperature behavior of all the transport coefficients arising in these two mechanisms are studied and presented. This paper generalize our previous work in \cite{Chen}.

One comment we would like to remind the readers is, in the case where the parity is spontaneously broken by $\theta$, our result doesn't violate the Coleman-Mermin-Wagner theorem which states that in 2+1 dimensions the global continuous symmetry cannot be spontaneous broken, otherwise the correlation of the massless Goldstone boson would have infrared divergence. Since the spontaneous broken parity in our model is a discrete symmetry, no Goldstone boson is created. Moreover, our result shows that $\widetilde{\sigma}$ and $\widetilde{\eta}_A/s$ go to zero at $T_c$ with the same mean field exponent as the order parameter, $(T_c -T)^{0.5}$. However, $\frac{\widetilde{\kappa}}{\rho/\mu}$ and $\widetilde{\xi}$ exhibit $(T_c-T)^{-1.5}$ behavior near $T_c$, which is novel and quite different from the usual systems. It would be interesting if these properties can be realized in the simple field theory models.

Our bulk model has a $U(1)$ gauge symmetry. It is interesting to generalize our analysis to the non-Abelian case, and construct dual hydrodynamics with non-Abelian gauge symmetry. In 2+1 dimensions, the Hall conductivity can arise from a bare Chern-Simons term in the underlying theory, or from the induced one due to the quantum loop correction. It is known that the coupling constant of the non-Abelian CS term at $T=0$ has to be quantized in order to maintain the gauge invariance. But in the finite temperature case, the coupling of the non-Abelian induced CS term is temperature dependent, and the gauge symmetry cannot be preserved for arbitrary temperature, unless all orders of quantum corrections are taken into account \cite{Dunne}. Therefore, non-perturbative methods are important in computing the Hall conductivity. Since we have used the gravity-hydrodynamics correspondence as an non-perturbative approach to derive the Hall conductivity, our analysis may shed new lights on the non-perturbative calculation of the Hall conductivity for the boundary non-Abelian gauge theory, provided the planar diagrams are all we need for its quantum corrections, due to the nature of AdS/CFT correspondence.

The equations (1.8a)$\sim$(1.8d) of \cite{yarom2} point out that the magnetic viscosity, curl viscosity, thermal Hall and heat Hall conductivities are not independent quantities; they satisfy four thermodynamic relations which can be derived from the constraint of positive entropy production. In the case where $\theta$ is sourceless on the boundary such that $\widetilde{\zeta}_A=0$, $\widetilde{\zeta}_B=0$, the remaining thermodynamic equation becomes
\beq
 \frac{\partial P_0}{\partial \epsilon_0} T \widetilde{\xi}
 + \frac{\partial P_0}{\partial \rho_0} \left(\widetilde{\sigma} - \frac{\widetilde{\kappa}}{T}\right)=0 .
\eeq
We have numerically checked that this relation is indeed satisfied in our model. This provides another supporting evidence that the bulk theory turns out to incorporate the second law of thermodynamics in the boundary on in AdS/CFT correspondence.

Finally, we would like to point out that, in 2+1 dimensional (or more generally, in odd dimensional) quantum field theory, 
there exists the parity anomaly \cite{Niemi} analogous to the even dimensional axial anomaly. As the case that the gauge anomaly induces new transport coefficients in the hydrodynamics \cite{son3}, we expect to see the effect of this odd dimensional parity anomaly in the hydrodynamic transport too. We are studying the holographic realization of this effect in detail, and will present our result in the near future.


\section*{Acknowledgments}
SHD wishes to thank CQUeST of Sogang University for warm hospitality during the preparation of this paper, and to thank Ren\'e Meyer for helpful discussions. The authors are grateful to Shi Pu for useful discussions. SHD's work are supported by the NSC grant NSC 100-2811-M-003-010. This project is supported in part by the NSC, NCTS of ROC, and the CASTS of NTU.


\appendix

\section{Calculation of the first order scalar mode}

The equations of motion of the our bulk theory are given in (\ref{EOM}). The general ansatz of black brane background with backreaction from the pseudo scalar field is (\ref{BG}). The standard gravity-hydrodynamics approach requires to perturb the bulk system by promoting $u^{\mu}, M, Q, A_{\mu}^{ext}$ to the slow-varying functions of the boundary coordinates $x^{\mu}$. Under the derivative expansions of the ansatz as in (\ref{fullans}) without taking any probe limit, the $O(\epsilon^0)$ ansatz satisfies the $O(\epsilon^0)$ equations (\ref{EOM}). The $O(\epsilon)$ part of the ansatz satisfies the $O(\epsilon)$ equations of motion which are given in this and the next section.

For the scalar modes, there four dynamical equations at $O(\epsilon)$, arising from $E_{rr}^{(1)}$, $H(r) M^v{}^{(1)}$, $\frac{r^2 f}{H} E_{rr}^{(1)} + E_{vr}^{(1)}$ and $KG^{(1)}$. By applying the zeroth order equations of motion which also imply their derivatives by $Q$ and $M$ vanish, these four first order equations take the form
\bea
&&\quad \frac{H}{r^4} \frac{d}{dr}\left(\frac{r^4 h'}{H}\right) - \theta'\,\varphi'
  = \frac{\lambda H}{2 r^4} \frac{d}{dr}\left(\frac{r^4 f' \theta'}{H^2}\right)(\partial_x \beta_y-\partial_y \beta_x)\, , \label{smode1}
\\
&& \,
\frac{1}{r^2}\frac{d}{dr}\left(\frac{r^2 a_v'}{H}\right) + \frac{2 A' h'}{H}
  = \lambda \frac{2A\theta'}{r^2}(\partial_x \beta_y-\partial_y \beta_x)
  + \frac{2 \lambda \,\theta'}{r^2} B^{(ext)}\, ,  \label{smode2}
\\
&& 
\frac{1}{r^2}  \frac{1}{dr}\left(\frac{r^3 k}{H}\right) + \frac{1}{2r^2} \frac{1}{dr} \left[\frac{1}{dr}\left(r^4 f\right)\, \frac{h}{H}\right] 
+ \frac{A'}{2H} a_v' - \frac{1}{2r^2} \frac{1}{dr}\left(\frac{r^4 f \theta'}{H} \varphi \right) 
+ \frac{r^2 f \theta'}{H} \varphi' \nn \\
&& \qquad\qquad 
= \frac{2}{r}\,\partial_i \beta_i + \frac{\lambda}{4H^2}\, r^2 f'^2\theta'\,(\partial_x \beta_y-\partial_y \beta_x)\, , \label{smode3}
\eea
\bea
&& 
-\frac{1}{r^2 H}\frac{d}{dr} \left(\frac{r^4 \theta' k}{H}\right) 
+ \frac{1}{r^2 H}\frac{d}{dr} \left(\frac{r^4 f \varphi'}{H}\right)
- \frac{d^2 V}{d\theta^2} \varphi 
 - \frac{2}{r^2 H} \frac{d}{dr} \left(\frac{r^4 f \theta'}{H} h\right)
 \nn \\
&& \qquad\qquad
= -\frac{\theta'}{H} \partial_i \beta_i 
+ \lambda \left[\frac{1}{4 r^2 H} \frac{d}{dr}\left(\frac{r^4 f'^2}{H^2}\right) - \frac{2AA'}{r^2 H}\right] (\partial_x \beta_y-\partial_y \beta_x)  \nn \\
&& \qquad\qquad\quad
-\frac{2}{r H}\left[\frac{d}{dr}\left(r\frac{\partial\theta}{\partial_M}\partial_v M\right) + \frac{d}{dr}\left(r\frac{\partial\theta}{\partial_Q}\partial_v Q\right)\right]
-\frac{2\lambda A'}{r^2 H} B^{(ext)} , \label{smode4}
\eea
where $'$ denotes the derivative by r. There are also two constraint equations due to $g^{rr} E_{vr}^{(1)} + g^{vr} E_{vv}^{(1)}$ and $M^r{}^{(1)}$ which describe the relations between $\partial_v M,\, \partial_v Q$ and $\partial_i \beta_i$:
\bea
\left(
\frac{r}{H^2} \frac{\partial f}{\partial M} - \frac{2\,rf}{H^3} \frac{\partial H}{\partial M}  
+ \frac{r^2 f}{2H^2} \frac{\partial \theta}{\partial M} \theta' 
\right) \partial_v M +\Bigg( M \to Q \Bigg) \partial_v Q 
&=& \frac{r^2 f'}{2 H^2} \partial_i \beta_i \, ,\label{smodecons1}
\\
\left(
\frac{1}{H^3} \frac{\partial H}{\partial M} A'
- \frac{1}{H^2} \frac{\partial A'}{\partial M}
\right) \partial_v M
+\Bigg( 
M \to Q
\Bigg) \partial_v Q &=& \frac{A'}{H^2} \partial_i \beta_i \, .
\label{smodecons2}
\eea

To proceed, we go into the probe limit
\beq
\theta \to \lambda \theta, \qquad \varphi \to \lambda \varphi, \qquad \lambda \to 0 \, . \nn
\eeq
Then it follows the solutions in Section \ref{scalarmode}.

\section{Calculation of the first order vector mode}

For the $O(\epsilon)$ vector mode perturbations, there are two dynamical equations arising from $E^{ri}{}^{(1)}$ and $r^2 M^i{}^{(1)}$:
\bea
-\frac{1}{2r^2} \frac{d}{dr} \left(\frac{r^4 j_i{}'}{H} \right)  
\!&+&\! \frac{A'}{2H} a_i' \nn \\
&=&  \frac{\partial_v \beta_i}{r}
- \Big[
  \frac{r^2}{2} \frac{d}{dr} \left(\frac{1}{r^2 H} \frac{\partial H}{\partial M} \right)
  +\frac{1}{2} \theta' \frac{\partial \theta}{\partial M}
\Big] \partial_i M  
-  \Big[ M \to Q \Big] \partial_i Q 
\nn \\
&& \;
+ \lambda \, \Big\{
  \frac{H}{2 r^6 f'} \frac{d}{dr} \left(\frac{r^7 f'^2}{H^3}\right) \frac{\partial \theta}{\partial M}
 -\frac{1}{4H} \frac{d}{dr} \left(\frac{r^2 f'}{H}\right) \frac{\partial \theta'}{\partial M} 
\nn \\ 
&& \hspace{1.2cm} 
 -\frac{1}{4 r^2} \frac{d}{dr} \left(\frac{r^4 f' \theta'}{H^3} \frac{\partial H}{\partial M} \right)
 +\frac{1}{4r^2} \frac{d}{dr} \left(\frac{r^4 \theta'}{H^2} \frac{\partial f'}{\partial M} \right)
  \Big\}\, \epsilon_{ij} \, \partial_j M  
\nn \\
&& \;
+ \lambda \, \Big\{ M \to Q \Big\}\, \epsilon_{ij} \, \partial_j Q 
+ \lambda
\left[\frac{1}{4r^2}\frac{d}{dr}\left(\frac{r^4 f' \theta'}{H^2}\right)\right] \epsilon_{ij}\, \partial_v \beta_j
\, , \label{vmode1}
\\
-\frac{1}{H} \frac{d}{dr}\left( \frac{r^2 f a_i{}'}{H}\right)
\!&+\!& \frac{r^2 A'}{H^2} j_i{}' \nn \\
&=& \frac{A'}{H} \partial_v \beta_i 
+ \frac{1}{H} \frac{\partial A'}{\partial M} \partial_i M
+ \frac{1}{H} \frac{\partial A'}{\partial Q} \partial_i Q
 \nn \\
&& \;
+ \lambda \Big\{
 \frac{2}{H}
\left(
\frac{\partial \theta}{\partial M} A' - \frac{\partial A}{\partial M} \theta'
\right) \Big\}\,
 \epsilon_{ij}\: \partial_j M 
+\lambda \Big\{ M \to Q \Big\}\, \epsilon_{ij}\: \partial_j Q \nn\\
&& \;
- \lambda\,\frac{2 A \theta'}{H} \epsilon_{ij}\:\partial_v \beta_j - \lambda \,\frac{2 \theta'}{H} \epsilon_{ij}\, F_{vi}^{ext}
\, \label{vmode2}
\eea
For both equations, the first line on the right hand side is parity-even, and the rest is parity-odd. After taking the probe limit, the leading order ($O(\lambda^0)$) terms contain only parity-even part, while the subleading ($O(\lambda^2)$) terms contain both parity-even and parity-odd contribution.

There is one constraint equation from $\frac{r^2 f}{H} E_{ri}^{(1)} + E_{vi}^{(1)}$: 
\bea
&&\Bigg\{
 - \frac{1}{2H} \frac{d}{dr} 
 \left(
 r^2 \frac{\partial f}{\partial M}
 \right)
+ \frac{1}{2H^2} \frac{d}{dr} 
 \left( r^2 f \right) \frac{\partial H}{\partial M}
 + \frac{A'}{2H} \frac{\partial A}{\partial M}
 - \frac{r^2 f \theta'}{2H} \frac{\partial \theta}{\partial M}
\Bigg\} \partial_i M \nn \\
&& \qquad\qquad \qquad 
+\Bigg\{ M \to Q\Bigg\} \partial_i Q
= \frac{1}{2H} (r^2 f' - A A')\: \partial_v \beta_i
- \frac{A'}{2H} F_{vi}^{ext} \, .
\label{vmodecons}
\eea

By taking the probe limit and making use of $f, H, A$ solutions, one can continue to the analysis in Section \ref{vectormode}. The leading order gives rise to the parity-even transport coefficients $\sigma$. The parity-even part of $O(\lambda^2)$ equations are higher order corrections to the leading order hydrodynamics. The parity-even part of $O(\lambda^2)$ gives rise to the parity violating coefficients.

\end{document}